\pdfoutput=1

\documentclass[11pt]{article}

\usepackage[final]{acl}

\usepackage{times}
\usepackage{latexsym}

\usepackage[T1]{fontenc}

\usepackage[utf8]{inputenc}

\usepackage{microtype}

\usepackage{inconsolata}

\usepackage{graphicx}

\title{LLM-Guided Semantic Relational Reasoning for Multimodal Intent Recognition}

\newcommand{\correspondingauthor}[1]{
  \begingroup
  \renewcommand\thefootnote{}
  \footnotetext{#1}
  \endgroup
}

\author{
 \textbf{Qianrui Zhou\textsuperscript{1}},
 \textbf{Hua Xu\textsuperscript{1*}},
 \textbf{Yifan Wang\textsuperscript{2,1}},
 \textbf{Xinzhi Dong\textsuperscript{1}},
\\
 \textbf{Hanlei Zhang\textsuperscript{1}}
\\
\\
 \textsuperscript{1}Department of Computer Science and Technology, Tsinghua University
 \\
 \textsuperscript{2}School of Information Science and Engineering, Hebei University of Science and Technology
 \\
 \href{mailto:zgr22@mails.tsinghua.edu.cn,}{zgr22@mails.tsinghua.edu.cn},
 \href{mailto:xuhua@tsinghua.edu.cn}{xuhua@tsinghua.edu.cn}
}

\usepackage{multirow}
\usepackage{graphicx}
\usepackage{array}
\usepackage{mdframed}
\usepackage{makecell}
\usepackage{amsmath}
\usepackage{amssymb}
\usepackage{booktabs}

\begin{document}
\maketitle

\correspondingauthor{* Hua Xu is the corresponding author.}

\begin{abstract}
Understanding human intents from multimodal signals is critical for analyzing human behaviors and enhancing human-machine interactions in real-world scenarios. However, existing methods exhibit limitations in their modality-level reliance, constraining relational reasoning over fine-grained semantics for complex intent understanding. This paper proposes a novel LLM-Guided Semantic Relational Reasoning (LGSRR) method, which harnesses the expansive knowledge of large language models (LLMs) to establish semantic foundations that boost smaller models' relational reasoning performance. Specifically, an LLM-based strategy is proposed to extract fine-grained semantics as guidance for subsequent reasoning,  driven by a shallow-to-deep Chain-of-Thought (CoT) that autonomously uncovers, describes, and ranks semantic cues by their importance without relying on manually defined priors. Besides, we formally model three fundamental types of semantic relations grounded in logical principles and analyze their nuanced interplay to enable more effective relational reasoning. Extensive experiments on multimodal intent and dialogue act recognition tasks demonstrate LGSRR's superiority over state-of-the-art methods, with consistent performance gains across diverse semantic understanding scenarios. The complete data and code are available at
\url{https://github.com/thuiar/LGSRR}.
\end{abstract}

\section{Introduction}
Multimodal intent recognition aims to utilize information from both natural language and other nonverbal modalities (e.g. video and audio) to enable machines to discern intents within real-world scenarios. It holds significant research importance and has broad applications in domains such as human-computer interaction \cite{xu2019towards}, chatbot \cite{fan2022building}, intelligent transportation system \cite{kaffash2021big}, medical diagnosis \cite{tiwari2022cnn,moon2022multimodal} and other human-robot interaction systems \cite{paul2022intent,mi2019object}.

Prior works \cite{zhang2022mintrec,zhang2024mintrec} have pioneered this area by introducing large-scale multimodal intent datasets, attracting increasing research attention. These studies also adapt fusion strategies from multimodal sentiment analysis to construct initial baselines \cite{hazarika2020misa,tsai2019multimodal,rahman2020integrating}, laying the foundation for subsequent advancements. Recently, numerous studies focus on extracting and aligning modality-level semantics to improve multimodal fusion. For instance, TCL-MAP \cite{zhou2024token} uses token-level contrastive learning and modality-aware prompts to enhance fusion between text and non-verbal modalities, while SDIF-DA \cite{huang2023sdif} applies a shallow-to-deep framework that aligns modalities before fusing them through a Transformer \cite{vaswani2017attention} layer. Beyond fusion mechanisms, several approaches investigate diverse perspectives to enhance intent understanding, such as leveraging global video context \cite{sun2024contextual}, reducing noise and redundancy in nonverbal streams \cite{zhu2024inmunet} and employing multi-task optimization \cite{zhang2024multimodal}.

Despite these significant advancements, two critical issues remain in multimodal intent recognition. First, existing approaches primarily emphasize coarse-grained and modality-level semantics, which introduces substantial redundancy and noise \cite{zhu2024inmunet}, resulting in a considerable gap between the extracted features and the true intent. Second, these methods generally model relationships between semantic concepts with basic fusion mechanisms, capturing only a limited subset of the complex reasoning relationships essential for accurate intent recognition. Consequently, there is a need for relational reasoning methods concentrating on fine-grained semantics for multimodal intent recognition, presenting two main challenges: (1) extracting fine-grained and intent-related semantics from diverse modalities, and (2) modeling complex reasoning relationships between these semantics. 

Given the aforementioned limitations, the demonstrated strengths of LLMs in semantic understanding tasks \cite{lai2024lisa,xu2024exploring} offer a promising solution for capturing fine-grained semantics. While \citet{xu-etal-2024-exploring} first leverages LLMs to extract commonsense knowledge, the approach remains constrained to modality-level cues and requires manually specified information. In this work, we aim to further unlock the potential of LLMs by enabling them to independently discover high-level multimodal semantic concepts and provide reasoning guidance. Moreover, due to the inherent complexity of semantic relations, we draw inspiration from classical logic, mapping basic operators (“or,” “and,” “not”) to their semantic counterparts (relative importance, complementarity, and inconsistency) to model intricate semantic interactions through their structured composition.

\begin{figure*}[t!]
	\centering  
	\includegraphics[scale=.42]{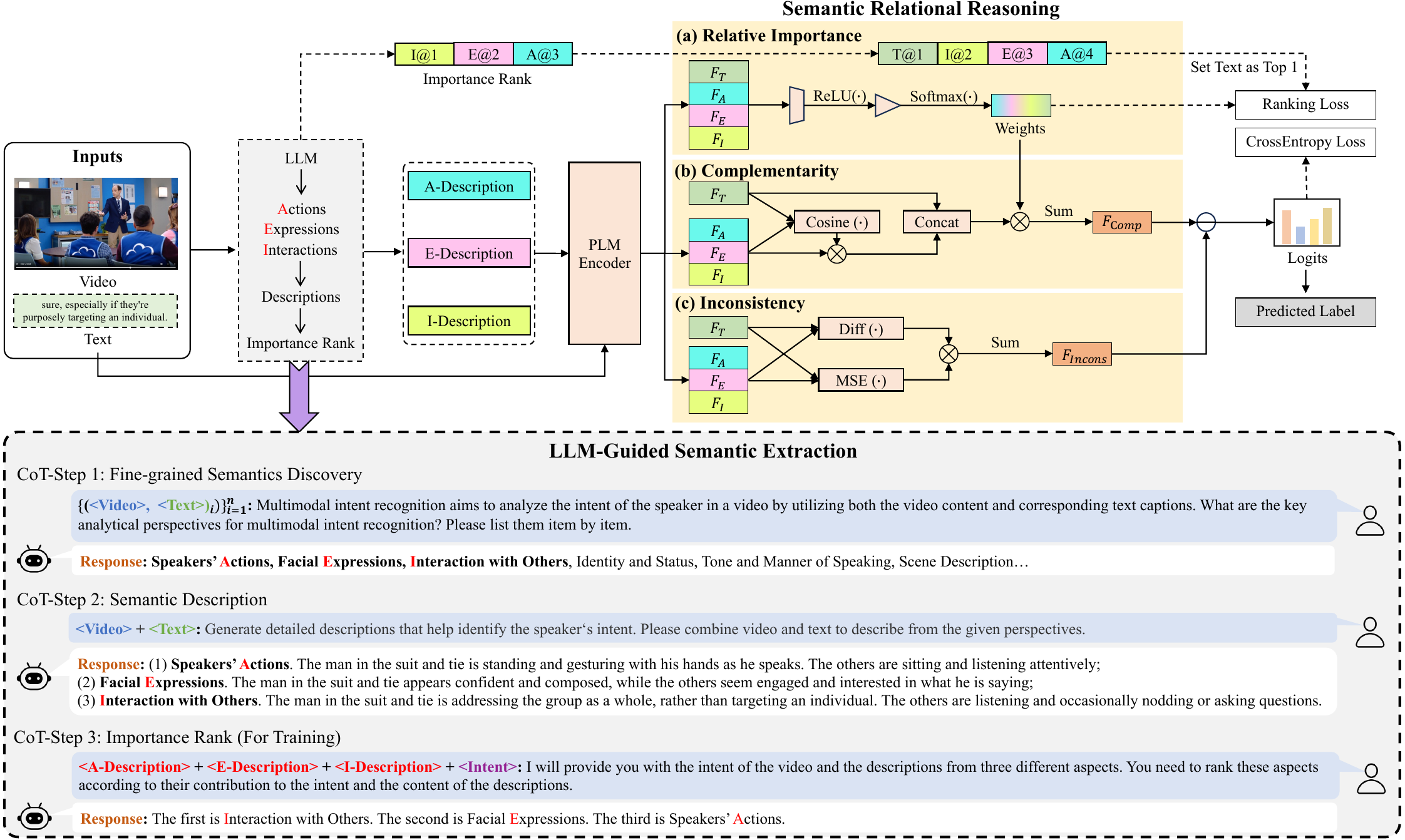}
	\caption{\label{model} Overall architecture of our proposed LLM-guided semantic relational reasoning (LGSRR) method.}
\end{figure*}

Consequently, we propose the LLM-Guided Semantic Relational Reasoning (LGSRR) framework, as illustrated in Figure~\ref{model}. To address the first challenge, we design an LLM-Guided Semantic Extraction module that employs a shallow-to-deep CoT for high-quality semantic discovery. Specifically, GPT-3.5 \cite{zhou2023comprehensive} is first prompted to identify fine-grained semantic cues relevant to multimodal intents, from which the top-$K$ most frequent cues are selected to conduct semantic extraction. These initial cues are then enriched by VideoLLaMA2 \cite{cheng2024videollama2}, which extracts detailed descriptive features from both text and video. Finally, GPT-3.5’s abductive reasoning capabilities are leveraged to generate a ranked semantic list, which serves as supervised guidance for subsequent relational reasoning. 
To tackle the second challenge, we introduce the Semantic Relational Reasoning module, which models the complex interplay among semantic cues by focusing on three fundamental logic-inspired relations. The relative importance of different semantics is learned through a unified weighting network optimized with NeuralNDCG \cite{pobrotyn2021neuralndcg} ranking loss based on the semantic rankings, while complementarity and inconsistency are naturally captured through cosine similarity and mean squared error between semantic representations respectively. To construct cohesive and discriminative intent representation, we integrate importance and complementarity as weighted factors to enrich the semantic space, while using inconsistency as a regularization term to ensure balanced reasoning.

Our contributions are summarized as follows: (1) We introduce an LLM-Guided framework which utilizes LLMs to autonomously acquire fine-grained semantics and deliver effective supervision for reasoning. To the best of our knowledge, this is the first attempt to employ LLMs to guide the learning of reasoning networks for multimodal intent recognition. (2) We propose a Semantic Relational Reasoning module that formally establishes three logic-driven relations and captures dynamic interactions among nuanced semantics, enabling structured and interpretable enhancement of multimodal reasoning. (3) We conduct extensive experiments on two challenging datasets for multimodal intent and dialogue act recognition respectively, achieving state-of-the-art performance.

\section{Related Works}

\subsection{Multimodal Intent Recognition}

Multimodal intent recognition is crucial for understanding human behavior by integrating verbal and nonverbal cues to capture real-world intents. While early datasets focus on simple semantic tasks \cite{kruk2019integrating, saha2020towards}, MIntRec \cite{zhang2022mintrec} advances the field through its diverse, fine-grained collection of multimodal samples, setting the first benchmark for multimodal intent recognition. Expanding on this, MIntRec2.0 \cite{zhang2024mintrec} increases the data scale and label diversity, while establishing multimodal fusion techniques \cite{zadeh2017tensor, liu2018efficient, zadeh2018memory,hazarika2020misa,tsai2019multimodal,rahman2020integrating} as baselines, thereby offering a more robust foundation for advancing intent recognition in complex contexts.

Recently, specialized intent recognition models have emerged to tackle unique challenges in this area. TCL-MAP \cite{zhou2024token} focuses on fusion strategies, using token-level contrastive learning and modality-aware prompts to enhance semantic depth. SDIF-DA \cite{huang2023sdif} employs a shallow-to-deep interaction module to align modalities across various levels, yielding high-quality fusion features. Additionally, CAGC \cite{sun2024contextual} uses video context to address perception biases and reduce intent uncertainty from multimodal inconsistencies, while InMu-Net \cite{zhu2024inmunet} applies an information bottleneck and multi-objective optimization to filter noise and redundancy in nonverbal data. MIntOOD \cite{zhang2024multimodal} introduces multi-granularity optimized objectives and employs dynamic weight fusion to enhance the robustness of multimodal representation.

\subsection{Multimodal Large Language Models}
Multimodal large language models (MLLMs) build on the success of LLMs in natural language processing by extending their capabilities to multimodal understanding. Early MLLMs primarily focus on aligning nonverbal modalities with LLMs’ input space. For instance, Flamingo \cite{alayrac2022flamingo} introduces gated cross-attention to handle interleaved multimodal data, while BLIP-2 \cite{li2023blip2} uses a Q-Former to map visual representations for integration with LLMs. Advanced video-capable MLLMs \cite{li2024videochat,zhang2023videollama, ataallah2024minigpt4video} address the challenge of extracting essential information from extensive visual content. VideoLLaMA2 \cite{cheng2024videollama2} employs spatial-temporal convolutions to capture dynamic visual details, while Qwen2-VL \cite{wang2024qwen2vl} and LLaVA-Next-Video \cite{zhang2024llavanextvideo} implement techniques like dynamic resolution and linear scaling for improved handling of varied frame sizes and longer videos. Recent models further leverage reasoning strategies to enhance understanding. Techniques such as CoT prompting \cite{wei2022cot} improve complex reasoning, and models like HuggingGPT \cite{shen2023hugginggpt} and VideoAgent \cite{wang2024videoagent} introduce agent-based planning, allowing the LLM to select or retrieve relevant frames and expert models, enhancing multimodal comprehension.

\section{Methodology}
\subsection{LLM-Guided Semantic Extraction}
\label{LLM-Guided Semantic Extraction}
Given raw video $V$ and text $T$, our goal is to extract fine-grained semantic features that closely align with intents for nuanced multimodal reasoning. To this end, we design a shallow-to-deep CoT consisting of the following three progressive steps. Details of CoT design are shown in Appendix \ref{details_of_CoT}.

\noindent \textbf{CoT-Step 1: Fine-grained Semantics Discovery}. To determine essential fine-grained semantics, we begin by randomly selecting a subset of samples \( B = \{(T_i, V_i) \mid i \in \{1, 2, \dots, n\}\} \) as background knowledge, in which each sample pair consists of corresponding text $T_i$ and video $V_i$. We then design $\text{Template}_{1}$ to prompt GPT-3.5 \cite{zhou2023comprehensive} to analyze these samples within the task context, and autonomously uncover salient semantic aspects, yielding an initial set $S_{\text{init}}$ of fine-grained concepts such as Speakers’ Actions (A), Facial Expressions (E), Interactions with Others (I), Identity and Status (IS), Tone and Manner of Speaking (TMS), Scene Description (SD), among others.
\begin{equation}
    S_{\text{init}} = \text{GPT-3.5}(B; \text{Template}_1).
\end{equation} 
From the set, we select top-$K$ frequent cues to form the refined set $S$ for further analysis, where the choice of $K=3$ is supported by the detailed comparison in Appendix~\ref{selection_of_semantic_aspects}.

\noindent \textbf{CoT-Step 2: Semantic Description}. Building on the fine-grained semantic set $S = \{A, E, I\}$, we employ VideoLLaMA2 \cite{cheng2024videollama2} to extract semantic descriptions $D_M$ for each concept $M \in S$. To ensure both descriptive quality and conceptual relevance, we craft structured prompts $\text{Template}_2$ specifically tailored to each semantic.
\begin{equation}
    D = \text{VideoLLaMA2}(V, T; S; \text{Template}_2).
\end{equation}
Consequently, we obtain high-quality descriptions $D = \{D_{A}, D_{E}, D_{I}\}$ for intent-related semantics, serving as a rich semantic foundation

\noindent \textbf{CoT-Step 3: Importance Rank}. To further incorporate LLM guidance into reasoning, we analyze the connections among the fine-grained semantics $S$ derived from the previous steps. Specifically, to bridge the comprehension gap between LLM and lightweight reasoning model, GPT-3.5 is utilized to generate a generalizable ranking of semantic contributions based on the ground-truth label $y$, using structured instructions $\text{Template}_3$:
\begin{equation}
    S_{\text{rank}} = \text{GPT-3.5}(D; y; \text{Template}_3),
\end{equation}
where the label $y$ is used only during training, allowing GPT-3.5 to perform abductive reasoning and provide supervisory signals that guide the reasoning module in evaluating semantic significance. The ranking statistics are presented in Appendix \ref{statistics_on_ranking_results}.

\subsection{Semantic Relational Reasoning}
\label{Semantic Relational Reasoning}
To move beyond simple fusion, we extend three core logical operations (“or,” “and,” and “not”) to semantic level, forming relations of relative importance, complementarity, and inconsistency. These relations underpin our Semantic Relational Reasoning module, capturing complex interactions for nuanced multimodal intent understanding. We argue that these core relational structures effectively represent the intricate dynamics necessary for robust intent recognition, drawing from logical reasoning principles in which complex relational patterns are built on basic operations.

For feature extraction from the input text and fine-grained semantic descriptions, we use BERT \cite{devlin2019bert}, a pre-trained language model commonly applied in intent recognition \cite{zhang2022mintrec}. To unify the feature space, we encode the text and semantic descriptions separately with BERT, accounting for stylistic differences. Specifically, given the text $T$ and the concatenated semantic descriptions $[D_{A}, D_{E}, D_{I}]$, we obtain their corresponding token representations as follows:
\begin{equation}
    Z_{\text{text}} = \text{BERT}(T), \quad Z_{\text{desc}} = \text{BERT}(D),
\end{equation}
where $Z_{\text{text}}$ and $Z_{\text{desc}}$ are token embeddings with dimensions $\mathbb{R}^{(l_T+1) \times d_T}$ and  $\mathbb{R}^{(3 * l_D+1) \times d_T}$, respectively. After encoding, we separate embeddings for actions $Z_A$, expressions  $Z_E$, and interactions $Z_I$. To acquire reasoning features, we apply mean pooling over the token embeddings:
\begin{equation}
    F_T = \text{Mean-Pooling}(Z_{\text{text}}),
\end{equation}
\begin{equation}
    F_M = \text{Mean-Pooling}(Z_{M}), \quad M \in \{A, E, I\}.
\end{equation}

\noindent \textbf{Relative Importance}. To account for the “or” operation among semantic components and highlight varying contributions, we apply weighted importance scores, a common approach in multimodal intent recognition \cite{rahman2020integrating}. Due to the limited supervision for these scores, we employ LLM-derived rankings with a ranking loss, NeuralNDCG \cite{pobrotyn2021neuralndcg}, to capture each semantic element’s relative significance. Specifically, we use a unified weight network with two linear layers, ReLU, and Softmax activations to produce normalized importance scores $\alpha_{T,A,E,I}$ for each semantic feature:
\begin{equation}
    h_{\{T,A,E,I\}} = \text{ReLU}(W_{1}F_{\{T,A,E,I\}} + b_{1}),
\end{equation}
\begin{equation}
    \alpha_{\{T,A,E,I\}} = \text{Softmax}(W_{2}h_{\{T,A,E,I\}} + b_{2}),
\end{equation}
where $W_{1}$, $W_{2}$, $b_{1}$, and $b_{2}$ are parameters. Given the central role of text, we prioritize the text feature $F_T$, assigning it the top rank in $R = \{R_T, R_{M_1}, R_{M_2}, R_{M_3}\}$, where $M_i$ denotes other semantic features. We then apply NeuralNDCG loss to align learned importance scores with LLM-derived rankings, formalized as:
\begin{equation}
    \mathcal{L}_{\text{rank}} = \frac{1}{N_R}\sum_{j \in R} \text{scale}(\widehat{P})_j g(\alpha_j) d(j),
\end{equation}
where $N$ is the size of $R$,  $g(\alpha_j)$ is the gain function, $d(j)$ is the rank discount, and $\text{scale}(\widehat{P})_j$ is the softmax-based similarity matrix, with a detailed explanation available in Appendix \ref{NeuralNDCG}.

\noindent \textbf{Complementarity}. In the intricate landscape of multimodal interactions, semantic components inherently possess complementary features that actively reinforce and validate one another, playing a pivotal role in decoding composite meanings. To capture this, we extend the logical “and” operation to emphasize complementarity between text and other semantic features. Specifically, we calculate the cosine similarity between the text feature $F_T$ and each fine-grained feature $F_M$, capturing cross-modal complementarity for richer representations. The complementarity score $\beta_{T, M}$ is given by:
\begin{equation}
\beta_{T, M} = \frac{F_T \cdot F_M}{||F_T|| \cdot ||F_M||}.
\end{equation}
To integrate complementarity relationship within the semantic space, we weight fine-grained semantic feature $F_M$ by its complementarity score  $\beta_{T, M}$, resulting in an enhanced feature representation:
\begin{equation}
    C_M = \beta_{T, M} \cdot F_M.
\end{equation}
We then combine relative importance and complementarity by concatenating each complementarity-enhanced feature $C_M$  with  $F_T$  and applying weighted averaging using importance scores for an integrated representation:
\begin{equation}
F_{\text{Comp}} = \sum_{M} \alpha_{M} \cdot \text{Concat}(F_T, C_M).
\end{equation}

\noindent \textbf{Inconsistency}. Modeling inconsistency among semantics is vital to understanding complex multimodal intents. For example, text might imply the \textit{Inform} intent, while gestures or expressions suggest \textit{Joke}. To capture such nuances, we extend the logical “not” operation to identify inconsistencies, enabling the model to interpret conflicting signals. Aligned with complementarity, we examine $F_T$  and other semantic features $F_M$ , creating an inconsistency penalty by calculating their differences:
\begin{equation}
    I_M = F_T - F_M.
\end{equation}
On the relational side, we use mean squared error to quantify this divergence, yielding inconsistency score $\gamma_{T, M}$ to reflect the degree of misalignment:
\begin{equation}
\gamma_{T, M} = \frac{1}{d} \sum_{i=1}^{d} (F_{T}^{(i)} - F_{M}^{(i)})^2,
\end{equation}
where $d$ is the dimension of $F_M$. Finally, we obtain the combination by weighting each  $I_M$ with its corresponding score $\gamma_{T, M}$, yielding the overall penalty feature $F_{\text{Incons}}$:
\begin{equation}
F_{\text{Incons}} = \sum_{M} \gamma_{T, M} \cdot I_M.
\end{equation}
This composite penalty feature $F_{\text{Incons}}$ captures complex semantic contradictions, enabling the model to identify subtle inconsistencies within multimodal intents, thereby refining its interpretative accuracy in complex contexts.

\begin{table*}[t!]\scriptsize
\centering
\resizebox{16cm}{!}{
\begin{tabular}{@{\extracolsep{2pt}}l|cccccc|cccccc}
    \toprule
     \multirow{2}{*}{Methods}& \multicolumn{6}{c|}{MIntRec2.0} & \multicolumn{6}{c}{IEMOCAP-DA}\\
   
    &\makecell{ACC ($\uparrow$)} 
    & \makecell{F1 ($\uparrow$)} 
    & \makecell{P ($\uparrow$)} 
    & \makecell{R ($\uparrow$)} 
    & \makecell{WF1 ($\uparrow$)}
    & \makecell{WP ($\uparrow$)}
    &\makecell{ACC ($\uparrow$)} 
    & \makecell{F1 ($\uparrow$)} 
    & \makecell{P ($\uparrow$)} 
    & \makecell{R ($\uparrow$)} 
    & \makecell{WF1 ($\uparrow$)}
    & \makecell{WP ($\uparrow$)}\\
    \midrule
    MISA & 55.16 & 49.51 & 51.80 & 49.92 & 55.05 & 57.06 & 73.76 & 72.26 & 73.03 & \underline{72.51} & 73.59 & 73.87 \\
    MAG-BERT & 60.38 & \underline{54.74} & 57.51 & \underline{54.54} & \underline{59.61} & 60.00 & 74.25 & 72.07 & 73.18 & 72.33 & 74.03 & 74.28 \\
    MulT & \textbf{60.66} & 54.12 & \underline{58.02} & 53.77 & 59.55 & \underline{60.12} & 73.74 & 72.28 & 73.40 & 72.21 & 73.66 & 74.15\\
    TCL-MAP & 58.24 & 52.25 & 54.28 & 52.41 & 57.24 & 57.55 & 74.37 & \underline{72.63} & \underline{74.02} & 72.39 & 74.21 & 74.76 \\
    SDIF-DA & 58.06 & 51.95 & 53.17 & 52.16 & 57.47 & 57.85 & 74.19 & 72.34 & 73.76 & 72.39 & 74.04 & \underline{74.77} \\
    MIntOOD & 58.25 & 51.73 & 56.79 & 50.99 & 57.11 & 58.65 &  \underline{74.56} & 71.31 & 72.70 & 70.89 &  \underline{74.40} & 74.65 \\
    \midrule
    LGSRR & \underline{60.46} & \textbf{55.35} & \textbf{59.33} & \textbf{55.09} & \textbf{59.72} & \textbf{60.85} & \textbf{74.95} & \textbf{72.99} & \textbf{74.27} & \textbf{72.74} & \textbf{74.88} & \textbf{75.47} \\
    \bottomrule 
\end{tabular}}
\caption{\label{main_results} Main Results comparing LGSRR with baselines on the MIntRec2.0 and IEMOCAP-DA datasets.}
\end{table*} 

\subsection{Training Objective}
For classification, we calculate predicted output  $\hat{y}$ with the obtained features and apply cross-entropy loss to optimize the model under the supervision of multi-class labels: 
\begin{equation}
    \hat{y} = W(F_{\text{Comp}} - F_{\text{Incons}}) + b,
\end{equation}
\begin{equation}
    \mathcal{L}_{\text{cls}} = -\frac{1}{N}\sum_{i=1}^N\sum_{c \in \mathcal{Y}} y_{i}^c \log(\operatorname{Softmax}(\hat{y}_i)^c),
\end{equation}
where $N$ is the batch size and $\mathcal{Y} = \{0, 1, \cdots, K-1\}$ denotes the label set. The overall objective is
\begin{equation}
    \mathcal{L} = \mathcal{L}_{\text{cls}} + \lambda \mathcal{L}_{\text{rank}},
\end{equation}
where $\lambda$ denotes the weight parameter.

\section{Experiments}
\subsection{Experimental Setting}
\noindent \textbf{Datasets}. MIntRec2.0 is a large-scale dataset with 30 fine-grained intent labels spanning text, video, and audio modalities. We adopt the original split with 6,165 training samples, 1,106 for validation, and 2,033 for test. For dialogue act classification, IEMOCAP-DA offers 12 annotated labels across the same three modalities, with 6,590 training samples, 942 for validation, and 1,884 for test.

\noindent \textbf{Baselines}. We compare LGSRR with state-of-the-art methods for multimodal intent recognition and dialogue act classification: (1) MulT \cite{tsai2019multimodal} introduces directional cross-modal attention to model interactions between modalities without requiring strict alignment; (2) MISA \cite{hazarika2020misa} projects each modality into modality-specific and modality-invariant subspaces, followed by self-attention for effective fusion; (3) MAG-BERT \cite{rahman2020integrating} incorporates a multimodal adaptive gating mechanism  that adjusts the text representation in semantic space using offsets computed from nonverbal modalities; (4) TCL-MAP \cite{zhou2024token} leverages token-level contrastive learning to enhance the textual modality by integrating visual and acoustic information, thereby promoting semantic acquisition and multimodal integration; (5) SDIF-DA \cite{huang2023sdif}  utilizes a shallow-to-deep interaction module that aligns and fuses modalities at both shallow and deep levels, capturing relations on all granularities; (6) MIntOOD \cite{zhang2024multimodal} employs a weighted feature fusion network to effectively model multimodal representations across multiple optimization granularities.

\noindent \textbf{Evaluation Metrics}. We evaluate model performance using following metrics: accuracy (ACC), F1-score (F1), precision (P), recall (R), weighted F1-score (WF1), and weighted precision (WP), where higher values indicate better performance.

\begin{table*}[t!]\scriptsize
\centering
\resizebox{16cm}{!}{
\begin{tabular}{@{\extracolsep{2pt}}l|cccccc|cccccc}
    \toprule
     \multirow{2}{*}{Ablation}& \multicolumn{6}{c|}{MIntRec2.0} & \multicolumn{6}{c}{IEMOCAP-DA}\\
   
    &\makecell{ACC ($\uparrow$)} 
    & \makecell{F1 ($\uparrow$)} 
    & \makecell{P ($\uparrow$)} 
    & \makecell{R ($\uparrow$)} 
    & \makecell{WF1 ($\uparrow$)}
    & \makecell{WP ($\uparrow$)}
    &\makecell{ACC ($\uparrow$)} 
    & \makecell{F1 ($\uparrow$)} 
    & \makecell{P ($\uparrow$)} 
    & \makecell{R ($\uparrow$)} 
    & \makecell{WF1 ($\uparrow$)}
    & \makecell{WP ($\uparrow$)}\\
    \midrule
   w / o LGSE & 58.57 & 52.83 & 54.37 & 52.72 & 58.22 & 58.69 & 74.55 & 70.40 & 71.32 & 70.64 & 74.40 & 74.73 \\
   LGSE (VideoLLaMA) & \underline{60.27} & \underline{54.91} & \underline{57.07} & \underline{54.78} & \underline{59.56} & \underline{60.00} & 74.26 & 71.62 & 72.51 & 71.46 & 74.14 & 74.55 \\
    LGSE (Qwen2-VL) & 59.59 & 54.62 & 56.63 & 54.58 & 58.89 & 59.20 & 74.54 & \underline{72.19} & \underline{73.05} & \underline{72.43} & 74.42 & \underline{75.05} \\
    w / o $\mathcal{L}_\text{rank}$ & 58.55 & 53.30 & 55.43 & 53.23 & 58.07 & 58.98 & 73.69 & 71.50 & 72.38 & 71.63 & 73.54 & 74.02 \\
    w / o SRR & 60.04 & 54.31 & 55.98 & 54.47 & 59.34 & 59.82 & \underline{74.57} & 71.80 & 71.82 & \underline{72.43} & \underline{74.44} & 74.63 \\
    \midrule
    Full & \textbf{60.46} & \textbf{55.35} & \textbf{59.33} & \textbf{55.09} & \textbf{59.72} & \textbf{60.85} &  \textbf{74.95} & \textbf{72.99} & \textbf{74.27} & \textbf{72.74} & \textbf{74.88} & \textbf{75.47} \\
    \bottomrule 
\end{tabular}}
\caption{\label{ablation_results} Ablation studies on the MIntRec2.0 and IEMOCAP-DA datasets.}
\end{table*} 

\begin{table*}[t!]\scriptsize
\centering
\resizebox{16cm}{!}{
\begin{tabular}{@{\extracolsep{2pt}}l|cccccc|cccccc}
    \toprule
     \multirow{2}{*}{MLLMs}& \multicolumn{6}{c|}{MIntRec2.0} & \multicolumn{6}{c}{IEMOCAP-DA}\\
   
    &\makecell{ACC ($\uparrow$)} 
    & \makecell{F1 ($\uparrow$)} 
    & \makecell{P ($\uparrow$)} 
    & \makecell{R ($\uparrow$)} 
    & \makecell{WF1 ($\uparrow$)}
    & \makecell{WP ($\uparrow$)}
    &\makecell{ACC ($\uparrow$)} 
    & \makecell{F1 ($\uparrow$)} 
    & \makecell{P ($\uparrow$)} 
    & \makecell{R ($\uparrow$)} 
    & \makecell{WF1 ($\uparrow$)}
    & \makecell{WP ($\uparrow$)}\\
    \midrule
    Qwen2-VL & \underline{28.63} & 7.18 & 8.68 & 7.74 & \underline{28.94} & \underline{37.53} & \underline{23.83} & \underline{7.39} & 10.14 & 9.43 & \underline{20.10} & \underline{37.09} \\
    MiniCPM-o 2.6 & 17.46 & \underline{14.75} & \underline{27.32} & \underline{14.97} & 17.98 & 37.11 & 14.01 & 6.93 & \underline{10.65} & \underline{11.62} & 13.56 & 36.29 \\
    VideoLLaMA2 & 17.17 & 4.95 & 12.32 & 4.65 & 13.10 & 25.55 & 15.02 & 2.97 & 7.05 & 4.11 & 7.59 & 29.77 \\
    \midrule
    LGSRR & \textbf{60.46} & \textbf{55.35} & \textbf{59.33} & \textbf{55.09} & \textbf{59.72} & \textbf{60.85} & \textbf{74.95} & \textbf{72.99} & \textbf{74.27} & \textbf{72.74} & \textbf{74.88} & \textbf{75.47} \\
    \bottomrule 
\end{tabular}}
\caption{\label{comparison_infer} Comparison of LGSRR with MLLMs across both datasets, with MLLMs evaluated via direct inference.}
\end{table*} 

\noindent \textbf{Implementation details}. For the LLM-Guided Semantic Extraction module, We utilize GPT-3.5-turbo for the first and third step and VideoLLaMA2-7B-16F for the second step. For feature extraction, the sequence lengths $l_T$ and $l_D$ are set as follows: (30, 50) for MIntRec2.0 and (70, 50) for IEMOCAP-DA, with a feature dimension $d_T$ of 1024. The training process includes 100 epochs, with a batch size of 32. We employ the PyTorch library via HuggingFace \cite{wolf2020transformers} for the pre-trained BERT model, optimized using AdamW \cite{loshchilov2018decoupled} with learning rates of (6e-6, 1e-5) for the respective datasets. For consistency, all reported results are the averages of five runs, using random seeds from 0 to 4, conducted on NVIDIA Tesla V100-SXM2s. The training cost is discussed in Appendix \ref{training_cost}.

\subsection{Results}
\noindent \textbf{Results on the MIntRec2.0 Dataset}. As illustrated in Table~\ref{main_results}, our LGSRR consistently surpasses existing SOTA methods, achieving the highest performance across five key metrics. In particular, LGSRR improves precision by 1.31\% over the top baseline, attributed to its ability to capture fine-grained semantics and perform effective relational reasoning. Moreover, LGSRR demonstrates substantial gains across the F1, R, and WP metrics with respective enhancements of 0.61\%, 0.55\%, and 0.73\%, further validating the effectiveness and robustness of our approach. Although MulT achieves a comparable ACC score, LGSRR maintains a clear edge across all other metrics, underscoring its comprehensive superiority in semantic understanding. The experimental results strongly demonstrate that our method excels in recognizing complex intent within multimodal scenarios by effectively leveraging fine-grained semantic relations.

\noindent \textbf{Results on the IEMOCAP-DA Dataset}. To comprehensively assess LGSRR’s effectiveness in various multimodal semantic tasks, we conduct experiments on the IEMOCAP-DA dataset. As shown in Table~\ref{main_results}, LGSRR exceeds state-of-the-art methods across all six metrics, demonstrating its robust capacity for nuanced semantic understanding. Notably, LGSRR achieves substantial gains with improvements of 0.58\%, 0.67\%, and 0.70\% in ACC, WF1 and WP, respectively, which underscores its ability to capture ambiguous concepts like dialogue acts. These results affirm LGSRR’s strong generalizability and position it as a promising framework for advancing multimodal semantic understanding.

\subsection{Ablation Study}
We evaluate the effects of key components, including the LLM-Guided Semantic Extraction module (LGSE), the ranking loss ($\mathcal{L}_\text{rank}$) and the Semantic Relational Reasoning module (SRR). Additionally, we also compare the performance of employing different MLLMs (VideoLLaMA \cite{zhang2023videollama} and Qwen2-VL \cite{wang2024qwen2vl}) in LGSE. The results are shown in Table~\ref{ablation_results}, confirming the individual contributions of the modules to the overall performance. The ablation study on the semantic relations within SRR is provided in Appendix \ref{ablations_for_semantic_relations}.

First, by removing the LGSE module, we utilize coarse-grained modality-level features from text, video, and audio extracted via BERT, Swin Transformer \cite{liu2021swin}, and WavLM \cite{chen2022wavlm}, respectively, to perform relational reasoning. This leads to substantial declines in all metrics across both datasets, with F1, P, and R scores dropping by over 2\%, demonstrating the improved alignment of fine-grained semantics with intent by effectively narrowing the semantic gap. When employing different MLLMs in LGSE, we observe no significant performance drop across most metrics, demonstrating the generalizability of our approach. Besides, removing the ranking loss ($\mathcal{L}_\text{rank}$) causes noticeable declines across metrics, underscoring the effectiveness of LLM guidance in capturing semantic feature importance and enhancing relational reasoning. Finally, replacing the SRR module with a simple summation-based fusion retains competitive ACC scores, suggesting the robustness of other components. However, declines in other metrics emphasize the SRR module’s role in refining semantic interactions and enhancing feature synergy, affirming each component’s unique contribution to nuanced multimodal understanding.

\subsection{Comparison to Frozen MLLMs}

To substantiate the rationale and effectiveness of the proposed LGSRR framework, we benchmark it against leading MLLMs on the MIntRec2.0 and IEMOCAP-DA datasets. To maintain consistency with the MLLM in LGSRR, we evaluate Qwen2-VL, MiniCPM-O 2.6 \cite{yao2024minicpm}, and VideoLLaMA2 in their 7B configurations using zero-shot reasoning without fine-tuning.

As Table~\ref{comparison_infer} shows, the experimental results on both datasets underscore the remarkable superiority of our proposed method over the state-of-the-art MLLMs. On the MIntRec2.0 dataset, LGSRR achieves an impressive 30\% improvement over the best-performing MLLM on all metrics other than WP, demonstrating its exceptional ability to model complex multimodal semantics. Likewise, on the IEMOCAP-DA dataset, LGSRR consistently outperforms MLLMs by over 35\% across all evaluation metrics, further affirming its robustness and adaptability. These results highlight LGSRR as a groundbreaking advancement in multimodal semantic understanding, offering a more efficient, scalable, and generalizable solution compared to MLLM-based approaches.

\begin{table}[t!]
\centering
\resizebox{\columnwidth}{!}{
\begin{tabular}{@{\extracolsep{2pt}}l|cccccc}
    \toprule
    \multirow{2}{*}{MLLMs} & \multicolumn{6}{c}{IEMOCAP-DA} \\
    & \makecell{ACC ($\uparrow$)} & \makecell{F1 ($\uparrow$)} & \makecell{P ($\uparrow$)} & \makecell{R ($\uparrow$)} & \makecell{WF1 ($\uparrow$)} & \makecell{WP ($\uparrow$)}\\
    \midrule
    Qwen2-VL & 68.10 & 59.85 & 56.91 & 66.75 & 67.09 & 68.72 \\
    VideoLLaMA2 & 71.07 & 68.57 & 71.90 & 68.91 & 71.32 & 73.62 \\
    LGSRR & \textbf{74.95} & \textbf{72.99} & \textbf{74.27} & \textbf{72.74} & \textbf{74.88} & \textbf{75.47} \\
    \bottomrule 
\end{tabular}}
\caption{\label{comparison_sft} Comparison of LGSRR with MLLMs on IEMOCAP-DA, with MLLMs evaluated via SFT.}
\end{table}

\subsection{Comparison to Fine-tuned MLLMs}
To further evaluate the practicality of our proposed method, we compared LGSRR with two cutting-edge MLLMs under the supervised fine-tuning settings on IEMOCAP-DA, as shown in Table~\ref{comparison_sft}. 

The results indicate that our method achieves SOTA performance across all metrics, showcasing its strong comprehensive capabilities. Specifically, compared to VideoLLaMA2 and Qwen2-VL, LGSRR yielded performance improvements of 3.88\% to 6.85\% and 4.42\% to 13.14\% on ACC and F1, highlighting our method's prominent capability for modeling intricate semantic relations. Moreover, the performance gains of LGSRR on P and R metrics, ranging from 2.37\% to 17.36\% and 3.83\% to 5.99\%, underscore the superiority of LGSRR in capturing deep semantic associations. Meanwhile, the stable performance on WF1 and WP metrics confirms the robustness of LGSRR when faced with imbalanced datasets. Overall, our method not only delivers more competitive performance but also avoids the computational burden of supervised fine-tuning, providing strong support and valuable insights for the precise optimization of smaller models for vertical tasks.

\begin{table}[t!]
\centering
\resizebox{\columnwidth}{!}{
\begin{tabular}{@{\extracolsep{2pt}}l|cccccc}
    \toprule
    \multirow{2}{*}{Methods} & \multicolumn{6}{c}{MIntRec2.0} \\
    & \makecell{ACC ($\uparrow$)} & \makecell{F1 ($\uparrow$)} & \makecell{P ($\uparrow$)} & \makecell{R ($\uparrow$)} & \makecell{WF1 ($\uparrow$)} & \makecell{WP ($\uparrow$)}\\
    \midrule
    Or & 59.46 & 53.51 & 54.69 & 53.44 & 58.90 & 59.12 \\
    And & \underline{59.76} & 54.08 & 56.07 & 54.01 & 59.02 & 59.36 \\
    Not & 59.53 & \underline{54.58} & 55.98 & 54.46 & 59.04 & 59.31 \\
    Combination & 59.63 & 54.13 & \underline{56.77} & \underline{54.95} & \underline{59.06} & \underline{59.41} \\
    \midrule
    LGSRR & \textbf{60.46} & \textbf{55.35} & \textbf{59.33} & \textbf{55.09} & \textbf{59.72} & \textbf{60.85} \\
    \bottomrule 
\end{tabular}}
\caption{\label{comparison_relations} Comparison of LGSRR with classic relations on the MIntRec2.0 dataset.}
\end{table}

\begin{figure}[t!]
	\centering  
	\includegraphics[scale=.62]{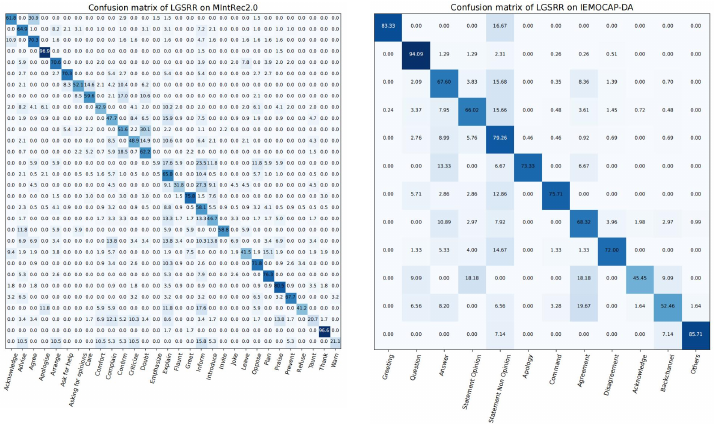}
	\caption{Confusion matrices on both datasets.}
    \label{confusion_matrices}
\end{figure}

\subsection{Comparison with Classic Relations}
To compare with classic relations, we define the corresponding feature-level operations and evaluate them on MIntRec2.0, as shown in ~\ref{comparison_relations}. Specifically, \textit{Or} is implemented by summing all semantic features, where the presence of any single salient cue can contribute effectively to intent recognition. \textit{And} is realized via the Hadamard product of semantic features, requiring all contributing cues to be jointly aligned with the target intent. \textit{Not} captures semantic inconsistency by computing pairwise feature differences, concatenating them, and projecting through a nonlinear layer. The \textit{Combination} setting integrates all three classic relation features by concatenation, followed by a nonlinear transformation to obtain a unified relational representation. 

The experimental results demonstrate the compelling superiority over classic relations across all metrics. Specifically, LGSRR outperforms the top classic relation method, achieving gains of 0.70\% in ACC and 0.77\% in F1, respectively. This reveals the enhanced feature representation capability of our semantic relationship reasoning module in complex multimodal scenarios. The significant 2.56\% performance increase on the P metric indicates that LGSRR is capable of identifying genuinely relevant semantic cues, thereby avoiding the misclassification of extraneous information as target intention features. Meanwhile, the model achieves performance improvements of 0.66\% and 1.44\% on WF1 and WP, respectively, which highlights that LGSRR can effectively alleviate the class imbalance problem. Overall, our semantic relational reasoning module outperforms classic relations methods, validating the efficacy of LGSRR in modeling fine-grained and intricate semantic interactions.

\subsection{Analysis of Classification Performance}
To investigate the fine-grained performance of classification, we plot the confusion matrices across all classes of our method on the MIntRec2.0 dataset and IEMOCAP-DA dataset, as shown in Figure~\ref{confusion_matrices}. Each score on the main diagonal represents the accuracy for the corresponding class.

On the MIntRec2.0 dataset, our method achieves over 70\% accuracy in 9 out of 30 classes. On the IEMOCAP-DA dataset, this ratio increases to 7 out of 12, highlighting LGSRR's superior performance. Specifically, LGSRR not only maintains high accuracy for simple intents like \textit{Apologize} and \textit{Thank}, but also excels in identifying complex intents like \textit{Arrange}, \textit{Ask for help} and \textit{Plan}, which require integrating multiple semantic cues. This demonstrates LGSRR’s capability to handle various levels of intent complexity. Limited performance on classes such as \textit{Joke} and \textit{Emphasize} is primarily due to less data in these categories. For dialogue act classification, LGSRR performs well across most categories, with only slight underperformance in more ambiguous classes such as \textit{Acknowledge} and \textit{Backchannel}. These results validate that LGSRR excels in distinguishing both fine-grained intents and broad communicative semantics through fine-grained semantic extraction and advanced relational reasoning.

\begin{figure*}[t!]
	\centering  
	\includegraphics[scale=.37]{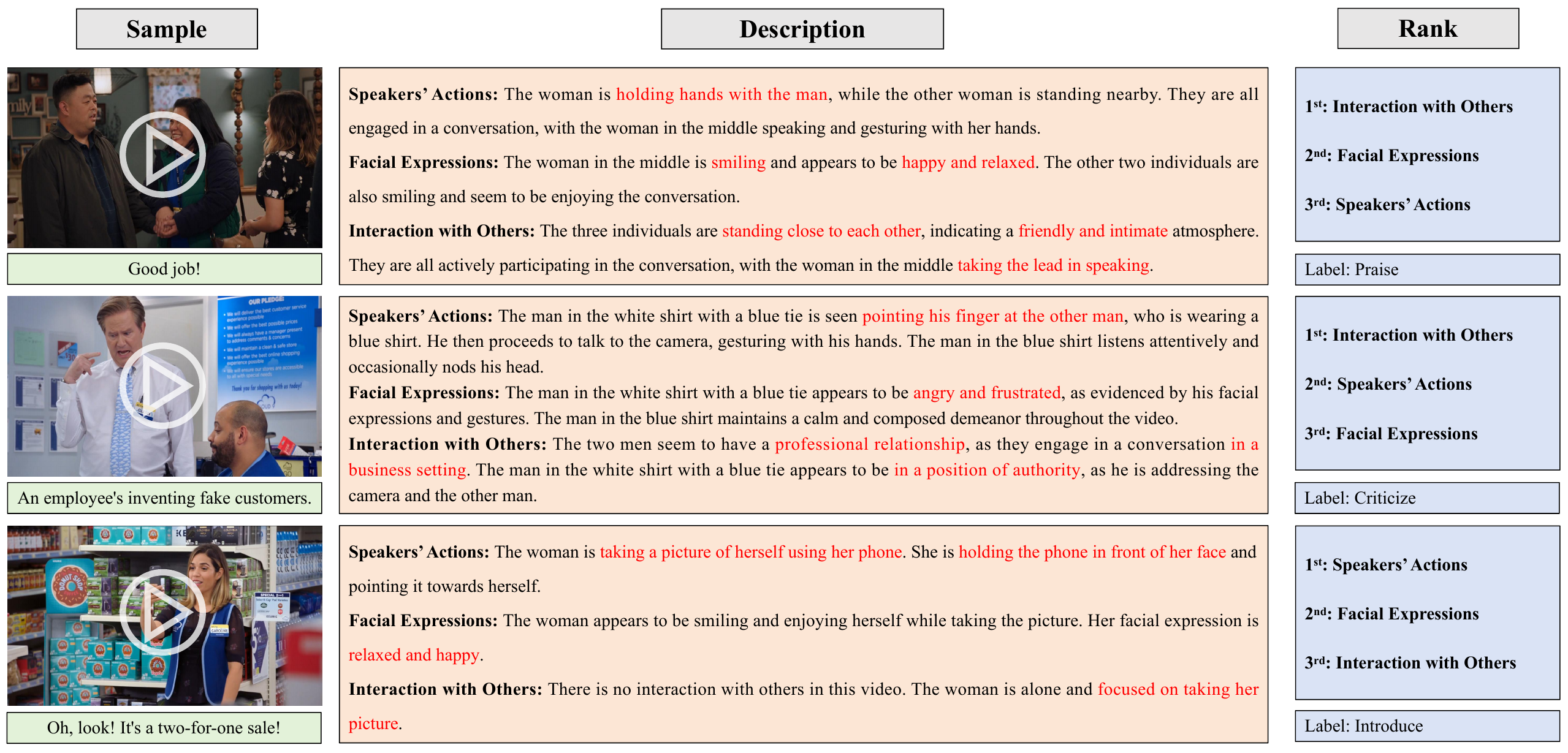}
	\caption{\label{Case_Study} Examples from the MIntRec2.0 dataset, showcasing descriptions and rankings of fine-grained semantics.
 }
\end{figure*}

\subsection{Case Study}
To offer deeper insight into semantic descriptions and rankings, we select three representative samples from the MIntRec2.0 dataset, covering diverse scenarios, multiple characters, varied emotions and distinct intents, as shown in Figure~\ref{Case_Study}. Each sample is presented with the raw data, detailed descriptions from three semantic perspectives and contribution rankings. Additional case studies from both datasets are provided in Appendix \ref{additional_case_studies}.

In terms of fine-grained semantic descriptions, our method accurately identifies significant details, including expressions like “smiling”, actions like “taking a picture of herself using her phone” and interactions like “standing close to each other”. Beyond basic cues, it also handles complex scenarios, such as ambiguous hand gestures and varying numbers of individuals, enabled by the effective CoT mechanism which significantly boosts the MLLM’s generative capabilities for nuanced semantics. For semantic ranking, our method leverages the LLM’s abductive reasoning ability to weigh the contribution of each fine-grained semantic relative to the true intent. In the first two examples, interpersonal interactions are prioritized consistent with the interaction concepts defined in the intent label \cite{zhang2024mintrec}, with actions and expressions ranked based on their direct relevance. In the third example, where a shopkeeper films a product introduction alone, the ranking correctly emphasizes the action, further demonstrating the strong capability of our semantic ranking approach.

\section{Conclusion}
In this paper, we present the LLM-Guided Semantic Relational Reasoning (LGSRR) framework to tackle the challenges of fine-grained semantic extraction and relational reasoning in multimodal intent recognition. Utilizing a shallow-to-deep CoT strategy, LGSRR harnesses the LLMs to autonomously uncover detailed semantics across modalities. By capturing logic-inspired relational patterns from logical, LGSRR effectively models intricate semantic relations to achieve superior representations. Our work not only demonstrates notable improvements across challenging multimodal classification tasks, but also carries significant implications for advancing LLM-guided frameworks in complex semantic understanding.

\section{Limitations}
This work has two primary limitations that warrant careful consideration. First, despite promising experimental results, the performance on these datasets still indicates substantial room for improvement, given the inherent complexity and variability of multimodal data. Second, although the study formally models basic semantic relations from a logical perspective, it does not fully account for the nuanced and context-dependent interactions in real-world scenarios. Future research could address these gaps by exploring more expressive relational structures or integrating adaptive reasoning mechanisms to better capture semantic complexity.


\section{Acknowledgements}
This work is supported by the National Natural Science Foundation of China (Grant No.62173195).

\bibliography{camera_ready}

\begin{thebibliography}{54}
\providecommand{\natexlab}[1]{#1}

\bibitem[{Alayrac et~al.(2022)Alayrac, Donahue, Luc, Miech, Barr, Hasson, Lenc, Mensch, Millican, Reynolds, Ring, Rutherford, Cabi, Han, Gong, Samangooei, Monteiro, Menick, Borgeaud, Brock, Nematzadeh, Sharifzadeh, Bi\'{n}kowski, Barreira, Vinyals, Zisserman, and Simonyan}]{alayrac2022flamingo}
Jean-Baptiste Alayrac, Jeff Donahue, Pauline Luc, Antoine Miech, Iain Barr, Yana Hasson, Karel Lenc, Arthur Mensch, Katherine Millican, Malcolm Reynolds, Roman Ring, Eliza Rutherford, Serkan Cabi, Tengda Han, Zhitao Gong, Sina Samangooei, Marianne Monteiro, Jacob~L Menick, Sebastian Borgeaud, and 8 others. 2022.
\newblock \href {https://proceedings.neurips.cc/paper_files/paper/2022/file/960a172bc7fbf0177ccccbb411a7d800-Paper-Conference.pdf} {Flamingo: a visual language model for few-shot learning}.
\newblock In \emph{Advances in Neural Information Processing Systems}, volume~35, pages 23716--23736. Curran Associates, Inc.

\bibitem[{Ataallah et~al.(2024)Ataallah, Shen, Abdelrahman, Sleiman, Zhu, Ding, and Elhoseiny}]{ataallah2024minigpt4video}
Kirolos Ataallah, Xiaoqian Shen, Eslam Abdelrahman, Essam Sleiman, Deyao Zhu, Jian Ding, and Mohamed Elhoseiny. 2024.
\newblock \href {https://arxiv.org/abs/2404.03413} {Minigpt4-video: Advancing multimodal llms for video understanding with interleaved visual-textual tokens}.
\newblock \emph{Preprint}, arXiv:2404.03413.

\bibitem[{Cadene et~al.(2019)Cadene, Ben-younes, Cord, and Thome}]{cadene2019murel}
Remi Cadene, Hedi Ben-younes, Matthieu Cord, and Nicolas Thome. 2019.
\newblock Murel: Multimodal relational reasoning for visual question answering.
\newblock In \emph{Proceedings of the IEEE/CVF Conference on Computer Vision and Pattern Recognition (CVPR)}.

\bibitem[{Chen et~al.(2022)Chen, Wang, Chen, Wu, Liu, Chen, Li, Kanda, Yoshioka, Xiao, Wu, Zhou, Ren, Qian, Qian, Wu, Zeng, Yu, and Wei}]{chen2022wavlm}
Sanyuan Chen, Chengyi Wang, Zhengyang Chen, Yu~Wu, Shujie Liu, Zhuo Chen, Jinyu Li, Naoyuki Kanda, Takuya Yoshioka, Xiong Xiao, Jian Wu, Long Zhou, Shuo Ren, Yanmin Qian, Yao Qian, Jian Wu, Michael Zeng, Xiangzhan Yu, and Furu Wei. 2022.
\newblock \href {https://doi.org/10.1109/JSTSP.2022.3188113} {Wavlm: Large-scale self-supervised pre-training for full stack speech processing}.
\newblock \emph{IEEE Journal of Selected Topics in Signal Processing}, 16(6):1505--1518.

\bibitem[{Chen et~al.(2019)Chen, Rohrbach, Yan, Shuicheng, Feng, and Kalantidis}]{chen2019graph}
Yunpeng Chen, Marcus Rohrbach, Zhicheng Yan, Yan Shuicheng, Jiashi Feng, and Yannis Kalantidis. 2019.
\newblock Graph-based global reasoning networks.
\newblock In \emph{Proceedings of the IEEE/CVF Conference on Computer Vision and Pattern Recognition (CVPR)}.

\bibitem[{Cheng et~al.(2024)Cheng, Leng, Zhang, Xin, Li, Chen, Zhu, Zhang, Luo, Zhao, and Bing}]{cheng2024videollama2}
Zesen Cheng, Sicong Leng, Hang Zhang, Yifei Xin, Xin Li, Guanzheng Chen, Yongxin Zhu, Wenqi Zhang, Ziyang Luo, Deli Zhao, and Lidong Bing. 2024.
\newblock \href {https://arxiv.org/abs/2406.07476} {Videollama 2: Advancing spatial-temporal modeling and audio understanding in video-llms}.
\newblock \emph{Preprint}, arXiv:2406.07476.

\bibitem[{Devlin et~al.(2019)Devlin, Chang, Lee, and Toutanova}]{devlin2019bert}
Jacob Devlin, Ming-Wei Chang, Kenton Lee, and Kristina Toutanova. 2019.
\newblock \href {https://doi.org/10.18653/v1/N19-1423} {{BERT}: Pre-training of deep bidirectional transformers for language understanding}.
\newblock In \emph{Proceedings of the 2019 Conference of the North {A}merican Chapter of the Association for Computational Linguistics: Human Language Technologies, Volume 1 (Long and Short Papers)}, pages 4171--4186, Minneapolis, Minnesota. Association for Computational Linguistics.

\bibitem[{Fan et~al.(2022)Fan, Wang, He, and Hu}]{fan2022building}
Yan Fan, Chengyu Wang, Peng He, and Yunhua Hu. 2022.
\newblock \href {https://doi.org/10.1145/3488560.3502189} {Building multi-turn query interpreters for e-commercial chatbots with sparse-to-dense attentive modeling}.
\newblock In \emph{Proceedings of the Fifteenth ACM International Conference on Web Search and Data Mining}, WSDM '22, page 1577–1580, New York, NY, USA. Association for Computing Machinery.

\bibitem[{Ganz et~al.(2024)Ganz, Kittenplon, Aberdam, Ben~Avraham, Nuriel, Mazor, and Litman}]{ganz2024question}
Roy Ganz, Yair Kittenplon, Aviad Aberdam, Elad Ben~Avraham, Oren Nuriel, Shai Mazor, and Ron Litman. 2024.
\newblock Question aware vision transformer for multimodal reasoning.
\newblock In \emph{Proceedings of the IEEE/CVF Conference on Computer Vision and Pattern Recognition (CVPR)}, pages 13861--13871.

\bibitem[{Grover et~al.(2019)Grover, Wang, Zweig, and Ermon}]{grover2018stochastic}
Aditya Grover, Eric Wang, Aaron Zweig, and Stefano Ermon. 2019.
\newblock \href {https://openreview.net/forum?id=H1eSS3CcKX} {Stochastic optimization of sorting networks via continuous relaxations}.
\newblock In \emph{International Conference on Learning Representations}.

\bibitem[{Hazarika et~al.(2020)Hazarika, Zimmermann, and Poria}]{hazarika2020misa}
Devamanyu Hazarika, Roger Zimmermann, and Soujanya Poria. 2020.
\newblock \href {https://doi.org/10.1145/3394171.3413678} {Misa: Modality-invariant and -specific representations for multimodal sentiment analysis}.
\newblock In \emph{Proceedings of the 28th ACM International Conference on Multimedia}, MM '20, page 1122–1131, New York, NY, USA. Association for Computing Machinery.

\bibitem[{Huang et~al.(2024)Huang, Qin, Wang, Tu, and Xu}]{huang2023sdif}
Shijue Huang, Libo Qin, Bingbing Wang, Geng Tu, and Ruifeng Xu. 2024.
\newblock \href {https://doi.org/10.1109/ICASSP48485.2024.10446922} {Sdif-da: A shallow-to-deep interaction framework with data augmentation for multi-modal intent detection}.
\newblock In \emph{ICASSP 2024 - 2024 IEEE International Conference on Acoustics, Speech and Signal Processing (ICASSP)}, pages 10206--10210.

\bibitem[{J{\"a}rvelin and Kek{\"a}l{\"a}inen(2002)}]{jarvelin2002cumulated}
Kalervo J{\"a}rvelin and Jaana Kek{\"a}l{\"a}inen. 2002.
\newblock Cumulated gain-based evaluation of ir techniques.
\newblock \emph{ACM Transactions on Information Systems (TOIS)}, 20(4):422--446.

\bibitem[{Jiang and Ye(2023)}]{jiang2023cross}
Ding Jiang and Mang Ye. 2023.
\newblock Cross-modal implicit relation reasoning and aligning for text-to-image person retrieval.
\newblock In \emph{Proceedings of the IEEE/CVF Conference on Computer Vision and Pattern Recognition (CVPR)}, pages 2787--2797.

\bibitem[{Kaffash et~al.(2021)Kaffash, Nguyen, and Zhu}]{kaffash2021big}
Sepideh Kaffash, An~Truong Nguyen, and Joe Zhu. 2021.
\newblock \href {https://doi.org/10.1016/j.ijpe.2020.107868} {Big data algorithms and applications in intelligent transportation system: A review and bibliometric analysis}.
\newblock \emph{International Journal of Production Economics}, 231:107868.

\bibitem[{Kruk et~al.(2019)Kruk, Lubin, Sikka, Lin, Jurafsky, and Divakaran}]{kruk2019integrating}
Julia Kruk, Jonah Lubin, Karan Sikka, Xiao Lin, Dan Jurafsky, and Ajay Divakaran. 2019.
\newblock \href {https://doi.org/10.18653/v1/D19-1469} {Integrating text and image: Determining multimodal document intent in {I}nstagram posts}.
\newblock In \emph{Proceedings of the 2019 Conference on Empirical Methods in Natural Language Processing and the 9th International Joint Conference on Natural Language Processing (EMNLP-IJCNLP)}, pages 4622--4632, Hong Kong, China. Association for Computational Linguistics.

\bibitem[{Lai et~al.(2024)Lai, Tian, Chen, Li, Yuan, Liu, and Jia}]{lai2024lisa}
Xin Lai, Zhuotao Tian, Yukang Chen, Yanwei Li, Yuhui Yuan, Shu Liu, and Jiaya Jia. 2024.
\newblock Lisa: Reasoning segmentation via large language model.
\newblock In \emph{Proceedings of the IEEE/CVF Conference on Computer Vision and Pattern Recognition (CVPR)}, pages 9579--9589.

\bibitem[{Li et~al.(2023)Li, Li, Savarese, and Hoi}]{li2023blip2}
Junnan Li, Dongxu Li, Silvio Savarese, and Steven Hoi. 2023.
\newblock \href {https://proceedings.mlr.press/v202/li23q.html} {{BLIP}-2: Bootstrapping language-image pre-training with frozen image encoders and large language models}.
\newblock In \emph{Proceedings of the 40th International Conference on Machine Learning}, volume 202 of \emph{Proceedings of Machine Learning Research}, pages 19730--19742. PMLR.

\bibitem[{Li et~al.(2024)Li, He, Wang, Li, Wang, Luo, Wang, Wang, and Qiao}]{li2024videochat}
KunChang Li, Yinan He, Yi~Wang, Yizhuo Li, Wenhai Wang, Ping Luo, Yali Wang, Limin Wang, and Yu~Qiao. 2024.
\newblock \href {https://arxiv.org/abs/2305.06355} {Videochat: Chat-centric video understanding}.
\newblock \emph{Preprint}, arXiv:2305.06355.

\bibitem[{Liu et~al.(2023)Liu, Wang, Huang, Wang, and Xu}]{liu2023cigar}
Yabo Liu, Jinghua Wang, Chao Huang, Yaowei Wang, and Yong Xu. 2023.
\newblock Cigar: Cross-modality graph reasoning for domain adaptive object detection.
\newblock In \emph{Proceedings of the IEEE/CVF Conference on Computer Vision and Pattern Recognition (CVPR)}, pages 23776--23786.

\bibitem[{Liu et~al.(2021)Liu, Lin, Cao, Hu, Wei, Zhang, Lin, and Guo}]{liu2021swin}
Ze~Liu, Yutong Lin, Yue Cao, Han Hu, Yixuan Wei, Zheng Zhang, Stephen Lin, and Baining Guo. 2021.
\newblock \href {https://doi.org/10.1109/ICCV48922.2021.00986} {Swin transformer: Hierarchical vision transformer using shifted windows}.
\newblock In \emph{2021 IEEE/CVF International Conference on Computer Vision (ICCV)}, pages 9992--10002.

\bibitem[{Liu et~al.(2018)Liu, Shen, Lakshminarasimhan, Liang, Zadeh, and Morency}]{liu2018efficient}
Zhun Liu, Ying Shen, Varun~Bharadhwaj Lakshminarasimhan, Paul~Pu Liang, Amir Zadeh, and Louis-Philippe Morency. 2018.
\newblock Efficient low-rank multimodal fusion with modality-specific factors.
\newblock \emph{arXiv preprint arXiv:1806.00064}.

\bibitem[{Loshchilov and Hutter(2019)}]{loshchilov2018decoupled}
Ilya Loshchilov and Frank Hutter. 2019.
\newblock \href {https://openreview.net/forum?id=Bkg6RiCqY7} {Decoupled weight decay regularization}.
\newblock In \emph{International Conference on Learning Representations}.

\bibitem[{Mi et~al.(2019)Mi, Tang, Deng, Görner, and Zhang}]{mi2019object}
Jinpeng Mi, Song Tang, Zhen Deng, Michael Görner, and Jianwei Zhang. 2019.
\newblock \href {https://doi.org/10.1016/j.cogsys.2018.12.010} {Object affordance based multimodal fusion for natural human-robot interaction}.
\newblock \emph{Cogn. Syst. Res.}, 54:128--137.

\bibitem[{Moon et~al.(2022)Moon, Lee, Shin, Kim, and Choi}]{moon2022multimodal}
Jong~Hak Moon, Hyungyung Lee, Woncheol Shin, Young-Hak Kim, and Edward Choi. 2022.
\newblock \href {https://doi.org/10.1109/JBHI.2022.3207502} {Multi-modal understanding and generation for medical images and text via vision-language pre-training}.
\newblock \emph{IEEE Journal of Biomedical and Health Informatics}, 26(12):6070--6080.

\bibitem[{Nam et~al.(2017)Nam, Ha, and Kim}]{nam2017dual}
Hyeonseob Nam, Jung-Woo Ha, and Jeonghee Kim. 2017.
\newblock Dual attention networks for multimodal reasoning and matching.
\newblock In \emph{Proceedings of the IEEE Conference on Computer Vision and Pattern Recognition (CVPR)}.

\bibitem[{Paul et~al.(2022)Paul, Sintek, Këpuska, Silaghi, and Robertson}]{paul2022intent}
Sheuli Paul, Michael Sintek, Veton Këpuska, Marius Silaghi, and Liam Robertson. 2022.
\newblock \href {https://doi.org/10.1109/ICMLA55696.2022.00127} {Intent based multimodal speech and gesture fusion for human-robot communication in assembly situation}.
\newblock In \emph{2022 21st IEEE International Conference on Machine Learning and Applications (ICMLA)}, pages 760--763.

\bibitem[{Pobrotyn and Białobrzeski(2021)}]{pobrotyn2021neuralndcg}
Przemysław Pobrotyn and Radosław Białobrzeski. 2021.
\newblock \href {https://arxiv.org/abs/2102.07831} {Neuralndcg: Direct optimisation of a ranking metric via differentiable relaxation of sorting}.
\newblock \emph{Preprint}, arXiv:2102.07831.

\bibitem[{Rahman et~al.(2020)Rahman, Hasan, Lee, Bagher~Zadeh, Mao, Morency, and Hoque}]{rahman2020integrating}
Wasifur Rahman, Md~Kamrul Hasan, Sangwu Lee, AmirAli Bagher~Zadeh, Chengfeng Mao, Louis-Philippe Morency, and Ehsan Hoque. 2020.
\newblock \href {https://doi.org/10.18653/v1/2020.acl-main.214} {Integrating multimodal information in large pretrained transformers}.
\newblock In \emph{Proceedings of the 58th Annual Meeting of the Association for Computational Linguistics}, pages 2359--2369, Online. Association for Computational Linguistics.

\bibitem[{Saha et~al.(2020)Saha, Patra, Saha, and Bhattacharyya}]{saha2020towards}
Tulika Saha, Aditya Patra, Sriparna Saha, and Pushpak Bhattacharyya. 2020.
\newblock \href {https://doi.org/10.18653/v1/2020.acl-main.402} {Towards emotion-aided multi-modal dialogue act classification}.
\newblock In \emph{Proceedings of the 58th Annual Meeting of the Association for Computational Linguistics}, pages 4361--4372, Online. Association for Computational Linguistics.

\bibitem[{Shen et~al.(2023)Shen, Song, Tan, Li, Lu, and Zhuang}]{shen2023hugginggpt}
Yongliang Shen, Kaitao Song, Xu~Tan, Dongsheng Li, Weiming Lu, and Yueting Zhuang. 2023.
\newblock \href {https://proceedings.neurips.cc/paper_files/paper/2023/file/77c33e6a367922d003ff102ffb92b658-Paper-Conference.pdf} {Hugginggpt: Solving ai tasks with chatgpt and its friends in hugging face}.
\newblock In \emph{Advances in Neural Information Processing Systems}, volume~36, pages 38154--38180. Curran Associates, Inc.

\bibitem[{Sun et~al.(2024)Sun, Xie, Ye, and Zhang}]{sun2024contextual}
Kaili Sun, Zhiwen Xie, Mang Ye, and Huyin Zhang. 2024.
\newblock Contextual augmented global contrast for multimodal intent recognition.
\newblock In \emph{Proceedings of the IEEE/CVF Conference on Computer Vision and Pattern Recognition (CVPR)}, pages 26963--26973.

\bibitem[{Tiwari et~al.(2022)Tiwari, Pant, Elarabawy, Abd-Elnaby, Mohd, Dhiman, and Sharma}]{tiwari2022cnn}
Pallavi Tiwari, Bhaskar Pant, Mahmoud~M. Elarabawy, Mohammed Abd-Elnaby, Noor Mohd, Gaurav Dhiman, and Subhash Sharma. 2022.
\newblock \href {https://doi.org/10.1155/2022/1830010} {Cnn based multiclass brain tumor detection using medical imaging}.
\newblock \emph{Computational Intelligence and Neuroscience}, 2022(1):1830010.

\bibitem[{Tsai et~al.(2019)Tsai, Bai, Liang, Kolter, Morency, and Salakhutdinov}]{tsai2019multimodal}
Yao-Hung~Hubert Tsai, Shaojie Bai, Paul~Pu Liang, J.~Zico Kolter, Louis-Philippe Morency, and Ruslan Salakhutdinov. 2019.
\newblock \href {https://doi.org/10.18653/v1/P19-1656} {Multimodal transformer for unaligned multimodal language sequences}.
\newblock In \emph{Proceedings of the 57th Annual Meeting of the Association for Computational Linguistics}, pages 6558--6569, Florence, Italy. Association for Computational Linguistics.

\bibitem[{Vaswani et~al.(2017)Vaswani, Shazeer, Parmar, Uszkoreit, Jones, Gomez, Kaiser, and Polosukhin}]{vaswani2017attention}
Ashish Vaswani, Noam Shazeer, Niki Parmar, Jakob Uszkoreit, Llion Jones, Aidan~N Gomez, \L~ukasz Kaiser, and Illia Polosukhin. 2017.
\newblock \href {https://proceedings.neurips.cc/paper_files/paper/2017/file/3f5ee243547dee91fbd053c1c4a845aa-Paper.pdf} {Attention is all you need}.
\newblock In \emph{Advances in Neural Information Processing Systems}, volume~30. Curran Associates, Inc.

\bibitem[{Wang et~al.(2024{\natexlab{a}})Wang, Bai, Tan, Wang, Fan, Bai, Chen, Liu, Wang, Ge, Fan, Dang, Du, Ren, Men, Liu, Zhou, Zhou, and Lin}]{wang2024qwen2vl}
Peng Wang, Shuai Bai, Sinan Tan, Shijie Wang, Zhihao Fan, Jinze Bai, Keqin Chen, Xuejing Liu, Jialin Wang, Wenbin Ge, Yang Fan, Kai Dang, Mengfei Du, Xuancheng Ren, Rui Men, Dayiheng Liu, Chang Zhou, Jingren Zhou, and Junyang Lin. 2024{\natexlab{a}}.
\newblock Qwen2-vl: Enhancing vision-language model's perception of the world at any resolution.
\newblock \emph{arXiv preprint arXiv:2409.12191}.

\bibitem[{Wang et~al.(2024{\natexlab{b}})Wang, Zhang, Zohar, and Yeung-Levy}]{wang2024videoagent}
Xiaohan Wang, Yuhui Zhang, Orr Zohar, and Serena Yeung-Levy. 2024{\natexlab{b}}.
\newblock Videoagent: Long-form video understanding with large language model as agent.
\newblock \emph{European Conference on Computer Vision (ECCV)}.

\bibitem[{Wei et~al.(2022)Wei, Wang, Schuurmans, Bosma, ichter, Xia, Chi, Le, and Zhou}]{wei2022cot}
Jason Wei, Xuezhi Wang, Dale Schuurmans, Maarten Bosma, brian ichter, Fei Xia, Ed~Chi, Quoc~V Le, and Denny Zhou. 2022.
\newblock \href {https://proceedings.neurips.cc/paper_files/paper/2022/file/9d5609613524ecf4f15af0f7b31abca4-Paper-Conference.pdf} {Chain-of-thought prompting elicits reasoning in large language models}.
\newblock In \emph{Advances in Neural Information Processing Systems}, volume~35, pages 24824--24837. Curran Associates, Inc.

\bibitem[{Wolf et~al.(2020)Wolf, Debut, Sanh, Chaumond, Delangue, Moi, Cistac, Rault, Louf, Funtowicz, Davison, Shleifer, von Platen, Ma, Jernite, Plu, Xu, Le~Scao, Gugger, Drame, Lhoest, and Rush}]{wolf2020transformers}
Thomas Wolf, Lysandre Debut, Victor Sanh, Julien Chaumond, Clement Delangue, Anthony Moi, Pierric Cistac, Tim Rault, Remi Louf, Morgan Funtowicz, Joe Davison, Sam Shleifer, Patrick von Platen, Clara Ma, Yacine Jernite, Julien Plu, Canwen Xu, Teven Le~Scao, Sylvain Gugger, and 3 others. 2020.
\newblock \href {https://doi.org/10.18653/v1/2020.emnlp-demos.6} {Transformers: State-of-the-art natural language processing}.
\newblock In \emph{Proceedings of the 2020 Conference on Empirical Methods in Natural Language Processing: System Demonstrations}, pages 38--45, Online. Association for Computational Linguistics.

\bibitem[{Xu(2019)}]{xu2019towards}
Wei Xu. 2019.
\newblock \href {https://doi.org/10.1145/3328485} {Toward human-centered ai: a perspective from human-computer interaction}.
\newblock \emph{Interactions}, 26(4):42–46.

\bibitem[{Xu et~al.(2024{\natexlab{a}})Xu, Hua, Li, and Wang}]{xu2024exploring}
Yanzhi Xu, Yueying Hua, Shichen Li, and Zhongqing Wang. 2024{\natexlab{a}}.
\newblock \href {https://doi.org/10.18653/v1/2024.acl-long.6} {Exploring chain-of-thought for multi-modal metaphor detection}.
\newblock In \emph{Proceedings of the 62nd Annual Meeting of the Association for Computational Linguistics (Volume 1: Long Papers)}, pages 91--101, Bangkok, Thailand. Association for Computational Linguistics.

\bibitem[{Xu et~al.(2024{\natexlab{b}})Xu, Hua, Li, and Wang}]{xu-etal-2024-exploring}
Yanzhi Xu, Yueying Hua, Shichen Li, and Zhongqing Wang. 2024{\natexlab{b}}.
\newblock \href {https://doi.org/10.18653/v1/2024.acl-long.6} {Exploring chain-of-thought for multi-modal metaphor detection}.
\newblock In \emph{Proceedings of the 62nd Annual Meeting of the Association for Computational Linguistics (Volume 1: Long Papers)}, pages 91--101, Bangkok, Thailand. Association for Computational Linguistics.

\bibitem[{Yang et~al.(2023)Yang, Li, Wang, Lin, Azarnasab, Ahmed, Liu, Liu, Zeng, and Wang}]{yang2023mmreact}
Zhengyuan Yang, Linjie Li, Jianfeng Wang, Kevin Lin, Ehsan Azarnasab, Faisal Ahmed, Zicheng Liu, Ce~Liu, Michael Zeng, and Lijuan Wang. 2023.
\newblock \href {https://arxiv.org/abs/2303.11381} {Mm-react: Prompting chatgpt for multimodal reasoning and action}.
\newblock \emph{Preprint}, arXiv:2303.11381.

\bibitem[{Yao et~al.(2024)Yao, Yu, Zhang, Wang, Cui, Zhu, Cai, Li, Zhao, He et~al.}]{yao2024minicpm}
Yuan Yao, Tianyu Yu, Ao~Zhang, Chongyi Wang, Junbo Cui, Hongji Zhu, Tianchi Cai, Haoyu Li, Weilin Zhao, Zhihui He, and 1 others. 2024.
\newblock Minicpm-v: A gpt-4v level mllm on your phone.
\newblock \emph{arXiv preprint arXiv:2408.01800}.

\bibitem[{Zadeh et~al.(2017)Zadeh, Chen, Poria, Cambria, and Morency}]{zadeh2017tensor}
Amir Zadeh, Minghai Chen, Soujanya Poria, Erik Cambria, and Louis-Philippe Morency. 2017.
\newblock Tensor fusion network for multimodal sentiment analysis.
\newblock \emph{arXiv preprint arXiv:1707.07250}.

\bibitem[{Zadeh et~al.(2018)Zadeh, Liang, Mazumder, Poria, Cambria, and Morency}]{zadeh2018memory}
Amir Zadeh, Paul~Pu Liang, Navonil Mazumder, Soujanya Poria, Erik Cambria, and Louis-Philippe Morency. 2018.
\newblock Memory fusion network for multi-view sequential learning.
\newblock In \emph{Proc. AAAI Conf. Artif. Intell.}

\bibitem[{Zhang et~al.(2023)Zhang, Li, and Bing}]{zhang2023videollama}
Hang Zhang, Xin Li, and Lidong Bing. 2023.
\newblock \href {https://doi.org/10.18653/v1/2023.emnlp-demo.49} {Video-{LL}a{MA}: An instruction-tuned audio-visual language model for video understanding}.
\newblock In \emph{Proceedings of the 2023 Conference on Empirical Methods in Natural Language Processing: System Demonstrations}, pages 543--553, Singapore. Association for Computational Linguistics.

\bibitem[{Zhang et~al.(2024{\natexlab{a}})Zhang, Wang, Xu, Zhou, Gao, Su, jinyue Zhao, Li, and Chen}]{zhang2024mintrec}
Hanlei Zhang, Xin Wang, Hua Xu, Qianrui Zhou, Kai Gao, Jianhua Su, jinyue Zhao, Wenrui Li, and Yanting Chen. 2024{\natexlab{a}}.
\newblock \href {https://openreview.net/forum?id=nY9nITZQjc} {{MI}ntrec2.0: A large-scale benchmark dataset for multimodal intent recognition and out-of-scope detection in conversations}.
\newblock In \emph{The Twelfth International Conference on Learning Representations}.

\bibitem[{Zhang et~al.(2022)Zhang, Xu, Wang, Zhou, Zhao, and Teng}]{zhang2022mintrec}
Hanlei Zhang, Hua Xu, Xin Wang, Qianrui Zhou, Shaojie Zhao, and Jiayan Teng. 2022.
\newblock \href {https://doi.org/10.1145/3503161.3547906} {Mintrec: A new dataset for multimodal intent recognition}.
\newblock In \emph{Proceedings of the 30th ACM International Conference on Multimedia}, MM ’22, page 1688–1697. ACM.

\bibitem[{Zhang et~al.(2024{\natexlab{b}})Zhang, Zhou, Xu, Su, Evans, and Gao}]{zhang2024multimodal}
Hanlei Zhang, Qianrui Zhou, Hua Xu, Jianhua Su, Roberto Evans, and Kai Gao. 2024{\natexlab{b}}.
\newblock Multimodal classification and out-of-distribution detection for multimodal intent understanding.
\newblock \emph{arXiv preprint arXiv:2412.12453}.

\bibitem[{Zhang et~al.(2024{\natexlab{c}})Zhang, Li, Liu, Lee, Gui, Fu, Feng, Liu, and Li}]{zhang2024llavanextvideo}
Yuanhan Zhang, Bo~Li, haotian Liu, Yong~jae Lee, Liangke Gui, Di~Fu, Jiashi Feng, Ziwei Liu, and Chunyuan Li. 2024{\natexlab{c}}.
\newblock \href {https://llava-vl.github.io/blog/2024-04-30-llava-next-video/} {Llava-next: A strong zero-shot video understanding model}.

\bibitem[{Zhou et~al.(2023)Zhou, Li, Li, Yu, Liu, Wang, Zhang, Ji, Yan, He, Peng, Li, Wu, Liu, Xie, Xiong, Pei, Yu, and Sun}]{zhou2023comprehensive}
Ce~Zhou, Qian Li, Chen Li, Jun Yu, Yixin Liu, Guangjing Wang, Kai Zhang, Cheng Ji, Qiben Yan, Lifang He, Hao Peng, Jianxin Li, Jia Wu, Ziwei Liu, Pengtao Xie, Caiming Xiong, Jian Pei, Philip~S. Yu, and Lichao Sun. 2023.
\newblock \href {https://arxiv.org/abs/2302.09419} {A comprehensive survey on pretrained foundation models: A history from bert to chatgpt}.
\newblock \emph{Preprint}, arXiv:2302.09419.

\bibitem[{Zhou et~al.(2024)Zhou, Xu, Li, Zhang, Zhang, Wang, and Gao}]{zhou2024token}
Qianrui Zhou, Hua Xu, Hao Li, Hanlei Zhang, Xiaohan Zhang, Yifan Wang, and Kai Gao. 2024.
\newblock \href {https://doi.org/10.1609/aaai.v38i15.29656} {Token-level contrastive learning with modality-aware prompting for multimodal intent recognition}.
\newblock \emph{Proceedings of the AAAI Conference on Artificial Intelligence}, 38(15):17114--17122.

\bibitem[{Zhu et~al.(2024)Zhu, Cheng, Chen, Chen, Zhang, Wu, Zheng, and Xing}]{zhu2024inmunet}
Zhihong Zhu, Xuxin Cheng, Zhaorun Chen, Yuyan Chen, Yunyan Zhang, Xian Wu, Yefeng Zheng, and Bowen Xing. 2024.
\newblock \href {https://doi.org/10.1145/3664647.3681623} {Inmu-net: Advancing multi-modal intent detection via information bottleneck and multi-sensory processing}.
\newblock In \emph{Proceedings of the 32nd ACM International Conference on Multimedia}, MM '24, page 515–524, New York, NY, USA. Association for Computing Machinery.

\end{thebibliography}

\appendix

\section{Related Work for Multimodal Reasoning}
\label{related_work_for_multimodal_reasoning}
Multimodal reasoning serves as a cornerstone for advancing intelligent systems, empowering them to integrate and interpret information from diverse sources to tackle complex, real-world challenges. Early approaches primarily focus on attention-based mechanisms to extract essential features from multimodal data. For instance, DANs~\cite{nam2017dual} employs joint visual and textual attention to capture interactions between modalities. To further improve reasoning abilities, alignment between textual and nonverbal modalities becomes a focal point. MUREL~\cite{cadene2019murel} models intra-modal interactions through rich vectorial representations, and IRRA~\cite{jiang2023cross} leverages implicit fine-grained relation learning to align multimodal data effectively. Graph-based methods also gain traction for characterizing semantic information and relations across modalities. GloRe~\cite{chen2019graph} proposes a global reasoning framework to learn relationships between distant regions, and CIGAR~\cite{liu2023cigar} combines linguistic and visual knowledge to construct multimodal graph representations. Recently, the reasoning capabilities of large language models (LLMs) are explored to tackle multimodal tasks. MM-REACT~\cite{yang2023mmreact} utilizes ChatGPT to identify visual experts for task-specific solutions, while QA-ViT~\cite{ganz2024question} integrates query-related information into vision transformers to enhance reasoning capabilities. These advancements highlight the continuous evolution of multimodal reasoning techniques and their potential for addressing increasingly complex scenarios.

\section{Selection of Semantic Aspects}
\label{selection_of_semantic_aspects}

To validate the effectiveness and rationale of our selected semantics, we conduct experiments by progressively incorporating additional semantic aspects, following their occurrence frequency, into the three most frequent categories: Speaker’s Actions (A), Facial Expressions (E), and Interactions with Others (I). The newly introduced semantic aspects include Tone and Manner of Speaking (TMS), Identity and Status (IS), and Scene Description (SD). The results with additional semantic aspects, compared against those from the main experiment, are presented in Table~\ref{semantic_results}.

The experimental results clearly demonstrate that incorporating additional semantic aspects leads to a significant performance decline, reaffirming the validity of our selection strategy. Across both datasets, adding one, two, or all three new aspects results in a drop of over 1\% across nearly all evaluation metrics. This suggests that human intent is primarily conveyed through facial expressions, actions, and social interactions, whereas incorporating excessive semantic aspects introduces redundancy or less relevant information, thereby diminishing the model’s ability to focus on essential multimodal cues. Furthermore, results from the IEMOCAP-DA dataset provide deeper insights into the relative impact of different semantic aspects. The negative effects of TMS and IS are more pronounced than those of SD, likely because TMS and IS capture intrinsic speaker characteristics, which are challenging to infer from brief, single-turn conversations. In contrast, SD primarily characterizes the dialogue environment, posing minimal ambiguity but contributing less to intent recognition. These findings strongly confirm the adequacy of our selected semantic aspects, emphasizing the importance of carefully choosing those most relevant to multimodal intent recognition.

\begin{table*}[t!]\scriptsize
\centering
\resizebox{2\columnwidth}{!}{
\begin{tabular}{@{\extracolsep{2pt}}l|cccccc|cccccc}
    \toprule
     \multirow{2}{*}{Settings}& \multicolumn{6}{c|}{MIntRec2.0} & \multicolumn{6}{c}{IEMOCAP-DA}\\
   
    &\makecell{ACC ($\uparrow$)} 
    & \makecell{F1 ($\uparrow$)} 
    & \makecell{P ($\uparrow$)} 
    & \makecell{R ($\uparrow$)} 
    & \makecell{WF1 ($\uparrow$)}
    & \makecell{WP ($\uparrow$)}
    &\makecell{ACC ($\uparrow$)} 
    & \makecell{F1 ($\uparrow$)} 
    & \makecell{P ($\uparrow$)} 
    & \makecell{R ($\uparrow$)} 
    & \makecell{WF1 ($\uparrow$)}
    & \makecell{WP ($\uparrow$)}\\
    \midrule
    A+E+I+TMS & 58.99 & 53.61 & 55.62 & \underline{53.70} & \underline{58.27} & 58.79 & 73.41 & 70.41 & 72.37 & 69.66 & 73.27 & 73.69 \\
    A+E+I+TMS+IS & \underline{59.03} & 53.26 & 56.24 & 53.22 & \underline{58.27} & \underline{59.29} & 73.24 & 69.69 & 70.69 & 69.41 & 73.09 & 73.28 \\
    A+E+I+TMS+IS+SD & 59.00 & \underline{53.63} & \underline{56.29} & 53.25 & 58.25 & 58.94 & \underline{73.96} & \underline{70.46} & \underline{72.57} & \underline{69.71} & \underline{73.79} & \underline{74.33}\\
    \midrule
    LGSRR (A+E+I) & \textbf{60.46} & \textbf{55.35} & \textbf{59.33} & \textbf{55.09} & \textbf{59.72} & \textbf{60.85} & \textbf{74.95} & \textbf{72.99} & \textbf{74.27} & \textbf{72.74} & \textbf{74.88} & \textbf{75.47} \\
    \bottomrule 
\end{tabular}}
\caption{\label{semantic_results} Comparison of using different semantic aspects, where A, E, I, TMS, IS and SD represent Speaker's Actions, Facial Expressions, Interaction with Others, Tone and Manner of Speaking, Identity and Status, and Scene Description, respectively.}
\end{table*} 

\begin{table*}[t!]
    \centering
    \resizebox{1.4\columnwidth}{!}{
    \begin{tabular}{@{\extracolsep{10pt}}l|ccc|ccc}
        \toprule
        \multirow{2}{*}{Semantics}
            & \multicolumn{3}{c|}{MIntRec2.0} & \multicolumn{3}{c}{IEMOCAP-DA} \\
            
        & Rank@1 & Rank@2 & Rank@3 & Rank@1 & Rank@2 & Rank@3 \\ 		
        \midrule
        Speakers' Actions & 1,476 & 3,814 & 875 & 1,675 & 3,982 & 933 \\
        
        Facial Expressions & 523 & 1,450 & 4,192 & 604 & 1,542 & 4,444 \\   

        Interactions with Others & 4,166 & 901 & 1,098 & 4,311 & 1,066 & 1,213 \\
        \midrule
    \end{tabular}}
    \caption{\label{Rank}  The semantic ranking results of Speakers' Actions, Facial Expressions, and Interactions with Others on the MIntRec2.0 and IEMOCAP-DA Dataset.}
\end{table*} 

\section{Details of CoT}
\label{details_of_CoT}
This section details the progressive design of the CoT in LGSRR, providing a clearer illustration of the underlying reasoning mechanism.
For CoT-Step 1, the prompt includes an introduction to the task background along with raw data descriptions, enabling the LLM to develop a more comprehensive understanding of the target semantics, as shown in $\text{Template}_{1}$. It is then asked to list key analytical perspectives, prompting it to autonomously identify fine-grained semantic aspects that are crucial for multimodal understanding.
\begin{mdframed}[backgroundcolor=gray!10, linewidth=1pt, leftmargin=0.5mm, rightmargin=0.5mm, linecolor=black!80]
\textcolor{blue}{$\text{Template}_{1}$:}
Multimodal intent recognition aims to analyze the intent of the speaker in a video by utilizing both the video content and corresponding text captions. What are the key analytical perspectives for multimodal intent recognition? Please list them.
\end{mdframed}

For CoT-Step 2, we guide MLLMs to generate structured and detailed semantic descriptions across key semantic cues from the previous step, such as speakers’ actions, facial expressions, and interaction with others. This step ensures that the extracted semantics are not only modality-aware but also intent-relevant, laying a solid foundation for reasoning over semantic relationships. 
\begin{mdframed}[backgroundcolor=gray!10, linewidth=1pt, leftmargin=0.5mm, rightmargin=0.5mm, linecolor=black!80]
\textcolor{blue}{$\text{Template}_{2}$:}
Generate detailed descriptions that help identify the speaker's intent. Please combine video and text to describe from the following perspectives: (1) Speakers' Actions: $<$Action Instruction$>$; (2) Facial Expressions: $<$Expression Instruction$>$; (3) Interaction with Others: $<$Interaction Instruction$>$. Focus on these aspects to create a comprehensive description that would aid in recognizing the intentions behind the speakers' actions and words.
\end{mdframed}

To further incorporate LLM guidance in establishing reasoning relationships, we take fine-grained semantic descriptions as input and prompt the LLM to rank them according to the contribution of each semantic description for CoT-Step 3. This ranking provides explicit supervision signals for the reasoning module to learn semantic importance in a more interpretable and intent-aware manner.
\begin{mdframed}[backgroundcolor=gray!10, linewidth=1pt, leftmargin=0.5mm, rightmargin=0.5mm, linecolor=black!80]
\textcolor{blue}{$\text{Template}_{3}$:}
I will provide you with the intent of the video and descriptions of the video from three different semantics. You need to rank these semantics according to their contribution to the intent and the content of the descriptions. The three aspects are Speakers' Actions, Facial Expressions and Interaction with Others. The intent of the video is $<$Intent$>$. The descriptions are $<$Action Description$>$, $<$Expression Description$>$ and $<$Interaction Description$>$.
\end{mdframed}

\section{Statistics on Ranking Results}
\label{statistics_on_ranking_results}
To thoroughly evaluate the significance of different fine-grained semantics, we analyze the semantic ranking results on the MIntRec2.0 and IEMOCAP-DA datasets, as illustrated in Table~\ref{Rank}. The table summarizes the ranking distributions for Speakers’ Actions, Facial Expressions, and Interactions with Others across both datasets.

From the statistics, we observe a consistent trend across both datasets: Interactions with Others consistently has the highest number of Rank@1 samples, with counts of 4,166 and 4,311, followed by Speakers’ Actions with 1,476 and 1,675, while Facial Expressions ranks the lowest with 523 and 604. This ranking distribution reflects the interaction-centric nature of intent labels such as \textit{criticize} and \textit{question}, which are deeply rooted in social dynamics and algin with real-world intent distributions. Given the complexity of intent semantics, facial expressions generally serve as a coarse indicator of intents, whereas actions provide more decisive clues, as seen in intents like \textit{Leave} or \textit{Criticize}. When examining each fine-grained semantic, LGSRR effectively differentiates the importance of actions and expressions, as reflected in their distinct ranking distributions. For instance, in MIntRec2.0, the number of Rank@2 and Rank@3 samples for actions is 3,814 and 875, respectively, compared to 1,450 and 4,192 for expressions. These results underscore LGSRR’s nuanced understanding of semantic contributions and potential in handling complex multimodal semantic tasks.

\begin{table*}[t!]\scriptsize
\centering
\resizebox{2\columnwidth}{!}{
\begin{tabular}{@{\extracolsep{0pt}}l|cccccc|cccccc}
    \toprule
     \multirow{2}{*}{Ablations}& \multicolumn{6}{c|}{MIntRec2.0} & \multicolumn{6}{c}{IEMOCAP-DA}\\
   
    &\makecell{ACC ($\uparrow$)} 
    & \makecell{F1 ($\uparrow$)} 
    & \makecell{P ($\uparrow$)} 
    & \makecell{R ($\uparrow$)} 
    & \makecell{WF1 ($\uparrow$)}
    & \makecell{WP ($\uparrow$)}
    &\makecell{ACC ($\uparrow$)} 
    & \makecell{F1 ($\uparrow$)} 
    & \makecell{P ($\uparrow$)} 
    & \makecell{R ($\uparrow$)} 
    & \makecell{WF1 ($\uparrow$)}
    & \makecell{WP ($\uparrow$)}\\
    \midrule
   w / o Relative Importance & 58.76 & 52.99 & 54.11 & 52.84 & 58.21 & 58.31 & 74.65 & \underline{72.29} & \underline{73.41} & 72.02 & 74.56 & 74.91 \\
    w / o Complementarity & 58.80 & \underline{53.93} & 55.90 & 53.75 & 58.28 & 58.96 & 73.67 & 70.85 & 72.40 & 70.43 & 73.49 & 73.86 \\
    w / o Inconsistency & \underline{59.53} & 53.88 & \underline{56.39} & \underline{54.01} & \underline{58.72} & \underline{59.30} & \underline{74.81} & 71.78 & 72.26 & \underline{72.31} & \underline{74.65} & \underline{75.03} \\
    \midrule
    Full & \textbf{60.46} & \textbf{55.35} & \textbf{59.33} & \textbf{55.09} & \textbf{59.72} & \textbf{60.85} &  \textbf{74.95} & \textbf{72.99} & \textbf{74.27} & \textbf{72.74} & \textbf{74.88} & \textbf{75.47} \\
    \bottomrule 
\end{tabular}}
\caption{\label{reasoning_ablation_results} Ablation studies for the reasoning relations on the MIntRec2.0 and IEMOCAP-DA datasets, with each configuration presenting results for the exclusion of relative importance, complementarity, or inconsistency. Bold text denotes the best performance, while underlined text indicates the second-best.}
\end{table*} 

\begin{figure*}[t!]
	\centering  
	\includegraphics[scale=.37]{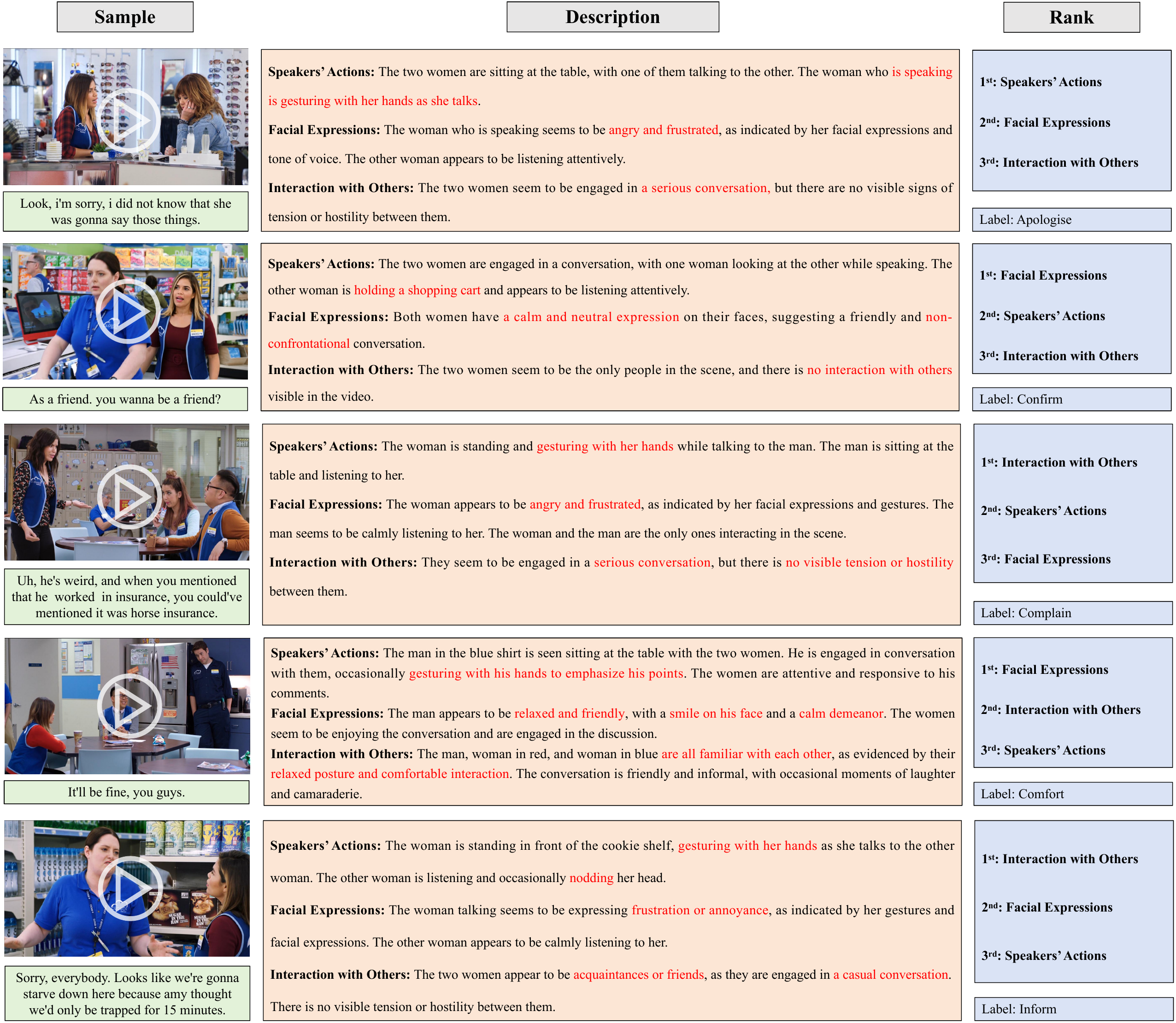}
	\caption{\label{Case_Study_MIntRec2} Samples from the MIntRec2.0, showcasing descriptions and ranking results of fine-grained semantics.
 }
\end{figure*}

\begin{figure*}[t!]
	\centering  
	\includegraphics[scale=.34]{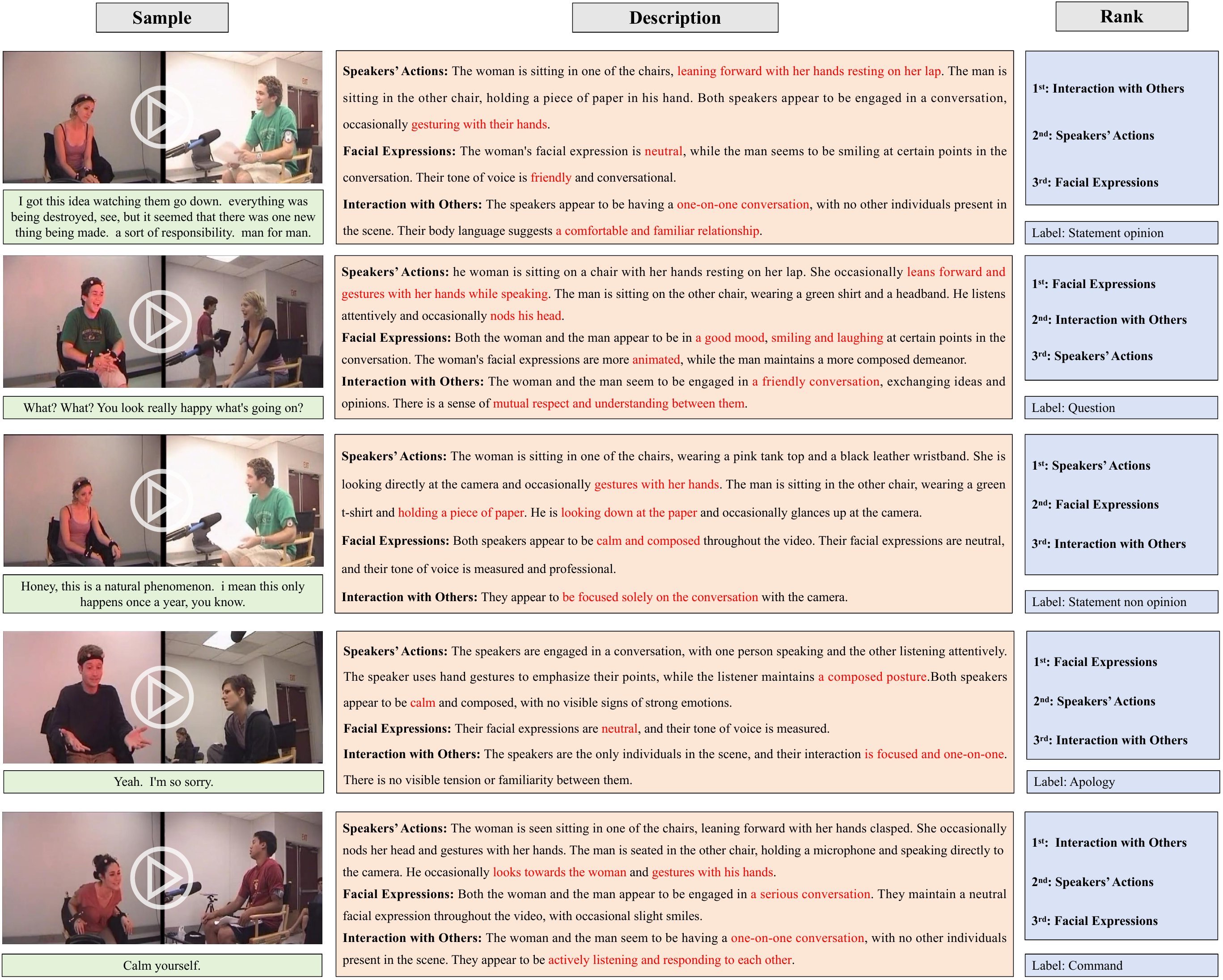}
	\caption{\label{Case_Study_IEMOCAP} Samples from the IEMOCAP-DA, showcasing descriptions and ranking results of fine-grained semantics.
 }
\end{figure*}

\section{Explanation of NeuralNDCG Loss}
\label{NeuralNDCG}
The NeuralNDCG loss \cite{pobrotyn2021neuralndcg} is a differentiable reformulation of the Normalized Discounted Cumulative Gain (NDCG) \cite{jarvelin2002cumulated} metric, tailored for learning-to-rank tasks by aligning model optimization directly with ranking performance. It approximates the sorting operation using a normalized soft permutation matrix $\widehat{P}$ and integrates gain and discount functions to compute ranking quality. The loss is defined as:
\begin{equation}
    \mathcal{L}_{\text{NeuralNDCG}} = \frac{1}{N_R} \sum_{j \in R} \text{scale}(\widehat{P})_j  g(s_j)  d(j),
\end{equation}
where $R$ represents the set of ranks, $N_R$ is its size, $j \in R$ corresponds to individual ranks, $s_j$ is the predicted score (interpreted as the importance score $\alpha_j$ in our work), $g(s_j) = 2^{s_j} - 1$ is the gain function that emphasizes the relevance of high-importance items, $d(j) = \frac{1}{\log_2(j + 1)}$ is the discount function that reduces the weight of lower-ranked elements, and $\text{scale}(\widehat{P})_j$ represents the row-stochastic approximation of the sorting operator. The matrix $\widehat{P}$ is derived by approximating the hard permutation matrix $P_{\text{sort}(s)}$, induced by sorting the predicted scores $s = f(x)$, with the formula~\cite{grover2018stochastic} as follows: 
\begin{equation}
    P = \frac{(n + 1 - 2u)s - A_s \mathbf{1}}{\tau},
\end{equation}
\begin{equation}
    \widehat{P}_{\text{sort}(s)}[u, :](\tau) = \text{softmax} \left( P\right),
\end{equation}
where $A_s[u, v] = |s_u - s_v|$ represents pairwise differences between scores, $\mathbf{1}$ is a column vector of ones, and $\tau > 0$ is the temperature parameter that controls the trade-off between the approximation accuracy and gradient stability. As $\tau \to 0$, $\widehat{P}$ converges to the true permutation matrix $P_{\text{sort}(s)}$, closely approximating the hard sorting process. To ensure both row- and column-stochasticity, Sinkhorn normalization is applied to $\widehat{P}$, further stabilizing its use in optimization by resolving inconsistencies in quasi-sorted outputs. NeuralNDCG integrates these components to enable gradient-based learning directly aligned with ranking performance, providing a powerful mechanism for optimizing ranking tasks in diverse applications.

\section{Training Cost}
\label{training_cost}
The additional computational cost primarily arises from the LLM-Guided Semantic Extraction process, which requires approximately two hours per dataset under the experimental conditions outlined. This duration is comparable to the training time of the model itself, which also takes around two hours. Despite this added computational expense, the trade-off is justified by the substantial benefits it brings in enhancing semantic understanding. Moreover, the training cost introduced by this approach remains considerably lower than that of fine-tuning, which typically demands 8–10 hours of computation. Thus, the proposed method achieves a more efficient balance between computational efficiency and semantic extraction performance.

\section{Ablations for Semantic Relations}
\label{ablations_for_semantic_relations}
To further validate the effectiveness of the three proposed reasoning relations, we conduct ablation studies on relative importance, complementarity, and inconsistency across the MIntRec2.0 \cite{zhang2024mintrec} and IEMOCAP-DA \cite{saha2020towards} datasets, with results summarized in Table~\ref{reasoning_ablation_results}. For relative importance and complementarity, we exclude their respective weight generation and weighting processes, while for inconsistency, the entire penalty feature is removed.

From the experimental results, the absence of relative importance leads to a notable drop in performance across all metrics on both datasets, highlighting its role as a cornerstone of our reasoning framework. On the MIntRec2.0 dataset, performance decreases range from 1.51\% to 5.22\%, which are significantly higher than the declines observed on IEMOCAP-DA, ranging from 0.30\% to 0.86\%. This emphasizes the critical importance of differentiating the relative significance of fine-grained semantics in understanding complex intents. Similarly, the removal of complementarity leads to metric reductions exceeding 1\% across the board, illustrating LGSRR’s success in capturing the inherent synergy between semantic elements. The absence of inconsistency results in the most significant declines in F1 and P scores, with reductions of 1.47\% and 2.94\% on MIntRec2.0 and 1.21\% and 2.01\% on IEMOCAP-DA. This decline is especially pronounced in complex intent categories such as \textit{Joke} and \textit{Flaunt}, which depend heavily on identifying contradictory semantic cues and are underrepresented in the dataset. These findings underscore LGSRR’s robustness in addressing nuanced inconsistencies, even in categories with limited data. These ablation studies confirm the essential role of the proposed reasoning relations and their integration, forming the basis of LGSRR’s capability to handle diverse intent semantics with precision and adaptability.

\section{Additional Case Studies}
\label{additional_case_studies}
Figure~\ref{Case_Study_MIntRec2} and Figure~\ref{Case_Study_IEMOCAP} present a diverse set of samples from the MIntRec2.0 and IEMOCAP-DA datasets, offering a detailed glimpse into the fine-grained semantic descriptions and importance rankings produced by LGSRR. The selected cases cover a range of intent labels and scenarios, each presenting semantic challenges and opportunities. On MIntRec2.0, LGSRR excels at identifying critical cues, such as “gesturing with hands” or “expressing frustration,” which are particularly relevant in contexts like \textit{Complain} and \textit{Criticize}. These results demonstrate the model’s ability to capture subtle yet impactful cues that are essential for intent understanding. In contrast, IEMOCAP-DA poses a greater challenge, featuring two-person dialogue scenes from varied perspectives that demand more nuanced reasoning. Fine-grained semantics in this dataset involve conversational dynamics rather than explicit physical actions, making it difficult to disentangle key semantic elements. For example, interactions are frequently characterized by subtle cues such as mutual respect, attentiveness, or slight gestures, which require precise modeling to capture effectively. Despite these complexities, LGSRR achieves impressive performance, consistently identifying the most relevant interactions and prioritizing critical semantic details. These case studies not only highlight the versatility and adaptability of LGSRR but also underscore its strength in navigating the intricacies of multimodal reasoning across datasets with distinct characteristics.
\end{document}